\documentclass[%
 aip, jcp,
 amsmath,amssymb,
 reprint,%
]{revtex4-2}

\usepackage{graphicx}
\usepackage{dcolumn}
\usepackage{bm}
\usepackage[utf8]{inputenc}    
\usepackage{ulem}
\usepackage{url}
\usepackage[dvipsnames]{xcolor}
\usepackage{hyperref}

\newcommand{\added}[1]{\textcolor{black}{#1}}

\begin{document}

\title{Plasmonic enhancement of molecular hydrogen dissociation on metallic magnesium nanoclusters}

\author{Oscar A. Douglas-Gallardo}
\affiliation{Department of Chemistry, University of Warwick, Gibbet Hill Road, Coventry, CV4 7AL, UK.}
\author{Connor L. Box}
\affiliation{Department of Chemistry, University of Warwick, Gibbet Hill Road, Coventry, CV4 7AL, UK.}
\author{Reinhard J. Maurer}
\email{r.maurer@warwick.ac.uk}
\affiliation{Department of Chemistry, University of Warwick, Gibbet Hill Road, Coventry, CV4 7AL, UK.}




\begin{abstract}
Light-driven plasmonic enhancement of chemical reactions on metal catalysts is a promising strategy to achieve highly selective and efficient chemical transformations. The study of plasmonic catalyst materials has traditionally focused on late transition metals such as Au, Ag, and Cu. In recent years, there has been increasing interest in the plasmonic properties of a set of earth-abundant elements such as Mg, which exhibit interesting hydrogenation chemistry with potential applications in hydrogen storage. This work explores the optical, electronic, and catalytic properties of a set of metallic Mg nanoclusters with up to 2057 atoms using time-dependent density functional tight-binding and density functional theory calculations. Our results show that Mg nanoclusters are able to produce highly energetic hot electrons with energies of up to 4~eV. By electronic structure analysis, we find that these hot electrons energetically align with electronic states of physisorbed molecular hydrogen, occupation of which by hot electrons can promote the hydrogen dissociation reaction.  We also find that the reverse reaction, hydrogen evolution \added{on metallic Mg}, can potentially be promoted by hot electrons, but following a different mechanism. Thus, from a theoretical perspective, Mg nanoclusters display very promising behaviour for their use in light promoted storage and release of hydrogen.

\end{abstract}
\maketitle

\section{Introduction}

Plasmonic properties arise when specific nanostructured materials interact with incident electromagnetic radiation.\cite{2000_elsayed,2003_elsayed,2006_elsayed,2011_hartland} This interaction produces a collective and coherent oscillation of conduction band  electrons, a phenomenon known as \textit{localized surface plasmon resonance} \added{(LSPR)}.\cite{2000_elsayed,2003_elsayed,2006_elsayed,2011_hartland} The associated optical absorption signal (\added{LSPR} band) is often broad and reaches a maximum when the incident light frequency ($\omega_{inc}$) is in resonance with the natural frequency of plasmonic excitation \added{($\omega_{LSPR}$)} of the irradiated nanomaterial.\cite{2000_elsayed,2003_elsayed,2006_elsayed,2011_hartland} This optical phenomenon leads to highly efficient light absorption  transferring large amounts of energy into the nanomaterial. The absorption cross-section associated with these plasmonic materials is often several orders of magnitude larger than most organic dye molecules.\cite{2000_elsayed,2003_elsayed,2006_elsayed,2019_douglas} Plasmonic nanomaterials have rapidly found different applications in several fields such as  optical-sensors,\cite{2015_optical-sensor} enhanced spectroscopies,\cite{2008_sers}  and even biomedical application.\cite{2010_dykman}

Recently, it has been reported that some plasmonic materials are able to effectively produce highly energetic distributions of electron-hole-pair (EHP) excitations upon non-radiative decay associated with dephasing of the plasmonic excitation (Landau damping).\cite{2014_clavero,2015_nordlander,2019_douglas,2020_berdakin,2019b_besteiro,2019_besteiro,2018_wei} 
EHPs act as hot-carriers and can potentially promote bond formation/dissociation events in molecules adsorbed on or nearby the plasmonic material. This is often discussed in terms of two possible mechanisms.\cite{2018_wei,2019_besteiro} EHPs created during the plasmon decay can lead to direct charge transfer into molecular states. This direct excitation mechanism depends strongly on the chemical interaction between the adsorbed molecule and the metal surface and is closely related to chemical interface damping (CID).\cite{2018_wei,2016_douglas,2010_negre,1993_persson,2017_link} CID arises from the hybridisation of adsorbate and metal states during molecular chemisorption and leads to a reduction of plasmonic excitation lifetime.\cite{2016_douglas,2010_negre,2018_wei} Highly excited non-equilibrium EHPs created upon plasmon decay will subsequently thermalize into an equilibrium distribution associated with an elevated electronic temperature that is higher than the lattice temperature.\cite{2000_elsayed,2011_hartland, 2013_rethfeld,2017_rethfeld,2019_besteiro} These equilibrated hot electrons can exchange energy indirectly with adsorbates. Finally, this electronic subsystem can subsequently transfer its energy to the lattice subsystem by means of electron-phonon scattering.\cite{2000_elsayed,2018_wei,2019_besteiro} This last stage is often modelled by a two-temperature model (TTM) which connects the electronic and lattice subsystem by means of two  non-linear coupled thermal diffusion equations.\cite{1974_anisimov} \added{LSPR} induced EHPs have shown to increase rates of important surface chemical reactions such as the CO$_2$ reduction,\cite{2018_jain,2017_halas} H$_2$O splitting \cite{2013_moskovits,2015_thomann,2012_moskovits} and H$_2$ dissociation reaction.\cite{2013_halas,2014_halas, 2016_halas}

Traditionally, noble metal (Au and Ag) nanostructures have been most widely studied as plasmonic catalyst materials primarily due to their chemical stability and characteristic optical response.\cite{2000_elsayed,2003_elsayed,2011_hartland,2006_elsayed,2010_dykman} Colloidal solutions of these plasmonic materials often exhibit a strong optical absorption band (\added{LSPR} band) which is located along the visible range of the electromagnetic spectrum.\cite{2011_hartland,2000_elsayed,2003_elsayed,2006_elsayed} The plasmonic excitation frequency can be widely tuned by modifying the nanostructure morphology and its chemical environment,\cite{2000_elsayed,2003_elsayed,2006_elsayed,2010_dykman,2011_hartland} but is typically limited to wavelengths larger than 400 nm for these specific noble plasmonic metals. In recent years the study of optical and electronic properties of non-standard plasmonic materials such as Al\cite{2008_schatz,2014_halas,2016_halas, 2017_douglas,2017_halas} and Mg\cite{2018_ringe,2015_sterl,2016_sterl,2020_ringe} has gathered increasing interest as they can provide highly cost effective alternatives.
Mg nanoclusters emerge as very attractive systems in the context of plasmonics  hot-electron effects.\cite{2018_ringe,2015_sterl,2020_ringe} Metallic Mg is an earth-abundant element that behaves as a pure plasmonic metal without sp to d interband transitions (like Au and Cu) due to lack of d-electrons in its electronic groundstate band structure. The plasmonic response for this type of  material is only due to sp-electrons and its dipole \added{LSPR} band can be easily detected even for relatively  small particle size. \cite{2004_solov,2021_sigalas}

From the point of view of catalysis, metallic Mg exhibits interesting hydrogenation chemistry with high reactivity towards forming hydride compounds (MgH$_2$).\cite{2006_johansson,2020_sterl,2018_sterl,2012_buckley,2018_ringe,2020_ringe}
Mg nanoparticles can switch from a metallic to an insulating state by simple uptake and release of molecular hydrogen.\cite{2020_sterl,2018_sterl,2016_sterl,2020_ringe} Magnesium can store a large amount of H atoms reaching up to 7.6~\% by weight as bulk material \added{representing  a promising, efficient and cost-effective alternative to store hydrogen molecules (chemical storage as metal hydride; chemisorption)}.\cite{1996_das, 2008_pozzo,2009_pozzo, 2012_buckley} \added{\cite{2005_jong,2015_zhang,2011_puru}} However, the activation energy associated with the H$_2$ dissociation reaction on metallic Mg surfaces (400 \AA{} thick Mg film) is relatively high (0.75$\pm$ 0.15) eV.\cite{2006_johansson,2009_pozzo,2013_lei} Different strategies have been explored to increase the effectiveness of this chemical reaction including doping with transition metals and the use of nanostructured particles.\cite{2006_du,2012_buckley,2011_urban,2011_prieto} \added{\cite{2005_jong,2015_zhang,2011_puru}} However, the potential effect of the Mg plasmonic activity on this particular chemical reaction has not been fully explored. A suitable hot-carrier distribution can potentially help to promote dissociative hydrogen adsorption on or release of molecular hydrogen from magnesium. 

We present a comprehensive computational study of the electronic and optical properties of a set of metallic Mg nanoclusters and periodic Mg(0001) slabs based on density functional theory (DFT) and time-dependent density functional tight-binding (TD-DFTB)\cite{2020_dftb+,2020_bonafe} calculations to simulate the optical absorption spectra and hot-electron distributions generated by incident light. We compare and contrast the optical properties of Mg with a conventional transition metal, namely Ag, before studying how hydrogen adsorption affects the plasmonic response of the nanoparticle. By analysis of the electronic structure and electron-nuclear response via  electronic friction theory,\cite{2016f_maurer, 2016_maurer,2017_maurer} we qualitatively assess the potential for plasmonic enhancement effects of dissociative hydrogen adsorption onto and hydrogen evolution from Mg nanoparticles and surfaces, which we discuss in the context of CID and Landau damping.

\section{Results and discussion}

\subsection{Optical absorption spectra and hot-carrier distribution on Mg nanoclusters}

\begin{figure}
 \centering
 \includegraphics[width=3.3in]{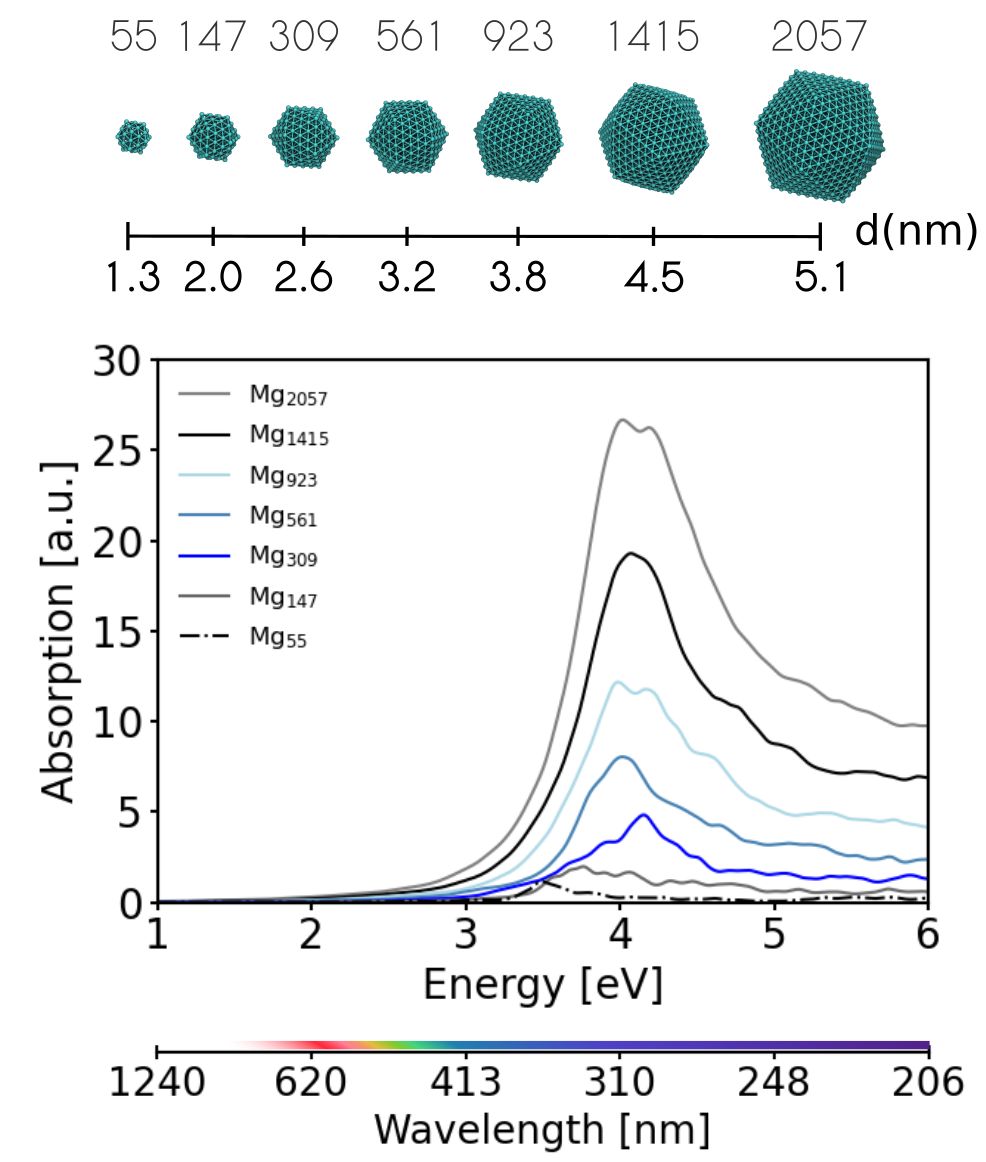}
 \caption{Schematic representation of the set of icosahedral Mg nanoclusters considered (labelled with the number of atoms)  along with their respective optical absorption spectrum. An extra x-axis was included to show the wavelength of the respective dipole \added{LSPR} energy in units of nanometers.}
 \label{fgr:mg-spectra}
\end{figure}

The optical absorption response and ensuing hot-carrier distribution are explored for a selected set of  icosahedral Mg nanoclusters\added{. The hot-carrier distribution is produced under} two different types of external electric fields, namely constant and pulsed irradiation \added{(\textit{cf.} section \ref{sec:methods})}. The number of metallic atoms contained in each particle follow the natural progression of icosahedral structures (with  magic numbers: 13, 55, 147, 309, 561, 923, 1415, 2057). The largest particle size that we have considered is $\sim$ 5 nm in diameter. \added{The most stable crystal structure for  Mg bulk is the hexagonal closest packed (hcp) configuration, however, for nanoscale sized particles, the icosahedral shape turns out to be the most stable geometric isomer.}\cite{1991_lange,2001_alhlrichs} Figure~\ref{fgr:mg-spectra} shows the optimized structures and the corresponding optical absorption spectra as calculated with the time-dependent density functional tight-binding (TD-DFTB) method (\textit{cf.} section~\ref{sec:methods} for computational details).


As can be seen in Figure~\ref{fgr:mg-spectra}, the main spectroscopic features of the nanoclusters are located within a range of 3-5 eV ($\sim$ 400-250 nm). The larger particles (561, 923, 1415 and 2057 atoms) are characterized by a strong single dipole \added{LSPR} band located at $\sim$ 4 eV ($\sim $ 310 nm). The \added{LSPR} band shows a strong dependence on particle size for the smaller particles, with the \added{LSPR} band peaks shifting from 3.5 to 4.5~eV as a function of nanocluster size. \added{For small particle sizes ranging from 55 to 309 atoms, we find a slight blue shift with increasing size. This may be related to a mixture of localised molecular and collective pseudo-plasmonic response. For larger nanocluster sizes, we observe a red shift, which is the expected behaviour for plasmonic response.} This optical behaviour is somewhat different from the case of Ag where only a red-shift is observed when the particle size is increased. \cite{2016_douglas} 
The excitation spectrum of the smallest Mg nanoparticle shows similarities with very small nanoclusters between 2-5 and $\sim$ 10-100 metallic atoms previously studied by correlated wavefunction theory\cite{2017_alok} and time-dependent DFT,~\cite{2004_solov,2021_sigalas} respectively. 

\begin{figure}[h]
 \centering
 \includegraphics[width=3.3in]{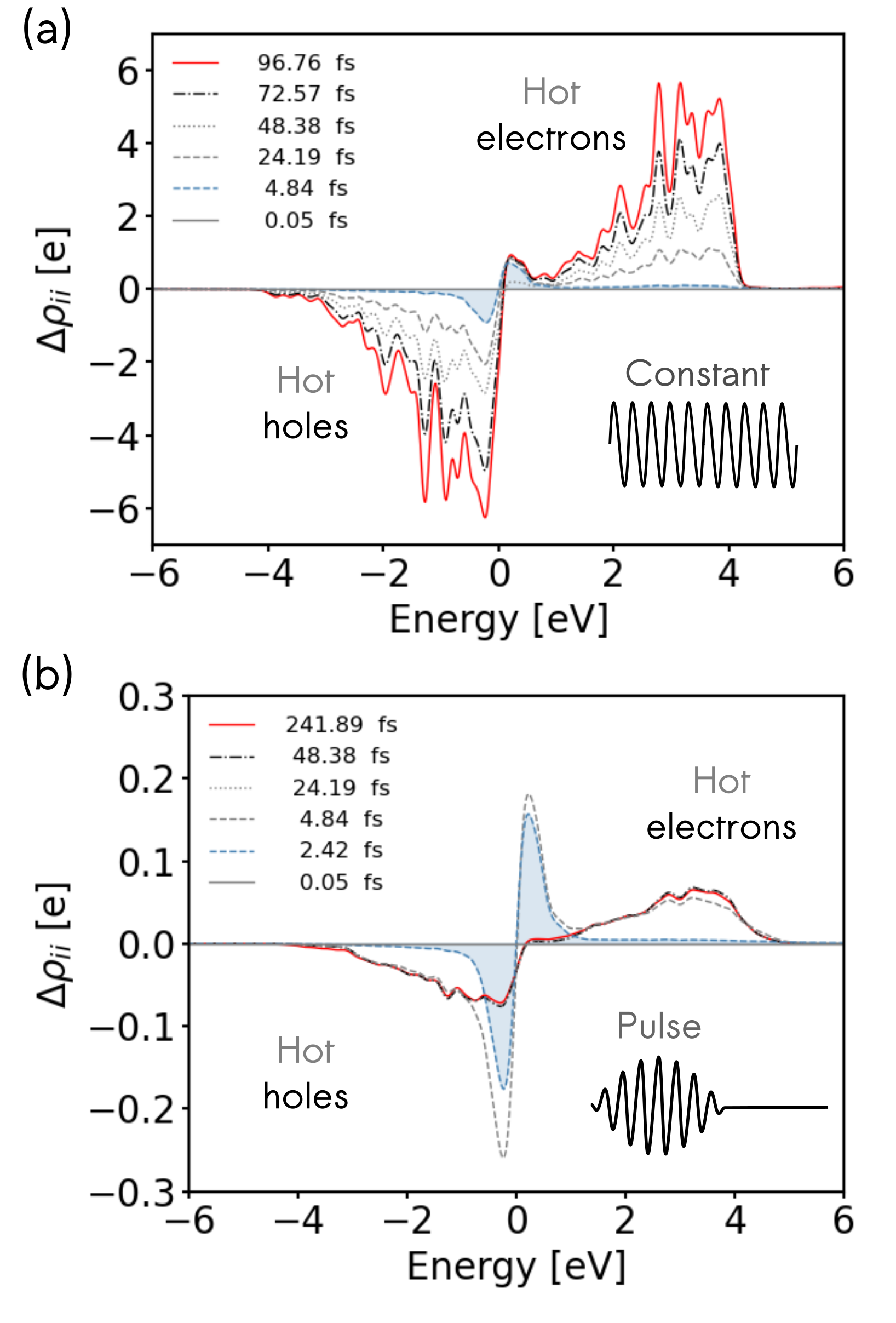}
 \caption{ Evolution of molecular orbital (MO) population at different time steps along the electronic dynamics with respect to the population at $t=0$. Two different types of external perturbation laser are used: (a) constant and (b) pulsed. An icosahedral Mg nanocluster with 1415 metallic atoms is illuminated with both sinusoidal time-dependent and sin$^2$ electric field tuned to the \added{LSPR} frequency (\added{$\omega_{LSPR}$} = 4.069 eV). The selected electric field intensity was 0.02 V $\AA^{-1}$.}
 \label{fgr:hot2d-mg}
\end{figure}

We further conduct an initial assessment of the EHP distribution of Mg produced under \added{external} laser irradiation (\added{pulsed and continuous excitation}). The underlying electronic dynamics associated with the plasmonic excitation, dephasing, and hot-carrier generation process are explored for a cluster with 1415 metallic atoms. Figure \ref{fgr:hot2d-mg} shows the EHP distribution that arises from the electronic dynamics following plasmonic excitation. A similar analysis has been conducted for noble metals in other recent reports.\cite{2016_douglas,2019_douglas,2020_berdakin} The population and depopulation dynamics of ground state molecular orbitals (MO) are distinguished with positive and negative population values, indicating hot electron and hot hole generation, respectively. These MO population differences are calculated with respect to the ground-state population at the initial time $t=0$ ($\Delta \rho_{ii}$=$\rho_{ii} (t)-\rho_{ii}(0)$). This is generated for both a continuous (panel a) and pulsed laser (panel b) source of external electric field at the \added{LSPR} frequency (\added{$\omega_{LSPR}$} = 4.069 eV). Sources were chosen in order to cover two extreme cases. In both cases, the same electric field intensity ($E_0$=0.02 V $\AA^{-1}$) was selected. The pulsed laser source used here is different from a Dirac $\delta$ pulse used to compute the optical absorption spectra in Fig.~\ref{fgr:mg-spectra} (\textit{cf.} section~\ref{sec:methods} for more details). For the pulsed source, the external electric field is modulated  using a sin$^2$ function with a laser duration of \added{$\tau=$}5~fs.

The electronic response of the Mg nanoparticle can be divided into two main stages. The first stage (highlighted in blue in Figure~\ref{fgr:hot2d-mg})  corresponds to plasmonic excitation located $\sim$ 1 eV above and below the Fermi energy and can be detected immediately after laser excitation has started. This particular time step is highlighted as it is close to the computed lifetime for plasmonic excitation associated with this Mg nanocluster (1.07 fs, \textit{vide infra}). The second stage involves EHP excitations that expand rapidly to higher energies above and below the Fermi level, producing a transient population and depopulation of a manifold of molecular states that evolves as a function of time. This last stage can be seen more clearly when a pulsed laser source is used. In this case, the finite energy intake leads to a convergence of the distribution as the TD-DFTB simulation does not fully capture thermalisation due to electron-electron scattering effects (\textit{vide infra)} and neglects equilibration with the lattice due to electron-phonon coupling. On the other hand, the continuous laser source steadily provides energy to the system and a continuous hot-carrier production is detected simultaneously with the plasmonic excitation. 

Plasmonic properties can be harnessed for catalysis in a continuous and pulsed illumination  regime.~\cite{qiDynamicControlElementary2020} In both cases, the largest concentration of high-energy electrons builds up at $\sim$3~eV above the Fermi energy and the largest concentration of holes is produced at $\sim$1~eV below the Fermi level. This particular distribution of EHPs can be connected to the Mg electronic bandstructure, which exhibits weakly dispersed valence bands close to the Fermi energy between M and L and at $\Gamma$ high symmetry points from which many electrons can be transferred into a high density of conduction bands at around 2 to 4~eV above the Fermi level (see Supporting Figure S1).

The non-equilibrium electronic distribution at different time steps of the dynamics for both external perturbations are shown on a logarithmic y axis scale in Supporting Figure S3. In this representation, equilibrium Fermi-Dirac distributions should appear as straight lines with the gradient governed by the effective electronic temperature. A quasi-equilibrium distribution can be found that ranges over few hundred meV above and below the Fermi level. Outside of this range, both illumination regimes generate highly non-equilibrium electron distributions that do not satisfy a Fermi-Dirac distribution. We find a characteristic steplike structure in the distribution with a step width of $\sim$ 4~eV. The non-equilibrium electronic distributions show the general features of ultrafast electron dynamics in metals reported from Boltzmann equation simulations.\cite{2013_rethfeld,2017_rethfeld,2000_elsayed} It is important to state that electronic thermalisation due to electron-electron scattering is not well described with the TD-DFTB method as it does not sufficiently capture the many-body correlation effects that give rise to such scattering events. For this reason, even long-time propagation for times beyond 200~fs do not show the electron thermalisation towards Fermi-Dirac distributions at elevated temperatures that is generally expected.\cite{2011_hartland,2019_besteiro,2018_wei,2016_atwater}
However, it is illustrative to generate a rough estimate of the electronic temperature regime that we expect to reach upon equilibration by computing the slope associated with the external edge of the steplike structure. From Supporting Figure S3, we estimate electronic temperature \added{ranges} of  9000$~\pm$\added{~3000~K} and  4400~$\pm$\added{~600~K} for constant and pulsed laser perturbation after about 100 and 240~fs, respectively. \added{These values provide an estimate of the range of electronic temperatures that can be expected upon equilibration.} We will subsequently use this in the discussion of plasmonic hydrogen evolution on Mg nanoparticles in section~\ref{sec:hydrogen}.

We have also explored the effect of particle size on hot-carrier generation.  Supporting Figure S4 reports the electronic dynamics after \added{4.84 and} 96.75~fs for Mg nanoclusters with 147, 309, 561, 923, 1415, and 2057 metallic atoms under constant illumination. All cluster sizes exhibit high energy hot electron generation around 4~eV. As we reduce the nanocluster size, the energy regime becomes more selective with a very pronounced peak between 3.5 and 4~eV for clusters with 147 and 309 atoms.  The generation of hot electrons within a narrow energy regime could potentially be very useful for the selective activation of chemical reactions and suggests a potentially measurable dependence of this effect with respect to nanocluster size. \added{Only clusters with 561 atoms or more exhibit the typical short-time plasmonic response of the EHP distribution close to the Fermi level. This plasmonic decay behaviour is fully established for larger clusters. Increasing the cluster size from 1415 to 2057 atoms does not further increase the population of hot electrons at 3.5 to 4~eV.} \added{This suggests that for Mg nanocluster sizes even larger than those  considered here (diameter $>$ 5 nm), the concentration of hot electrons may effectively decrease, which would affect the ability to activate surface chemical reactions by hot electron transfer. This size effect has been previously shown for Ag nanoparticles.\cite{2015_manjavacas}}

\subsection{Comparison of plasmonic response of Mg and Ag nanoclusters}

\begin{figure}[h]
 \centering
 \includegraphics[width=3.3in]{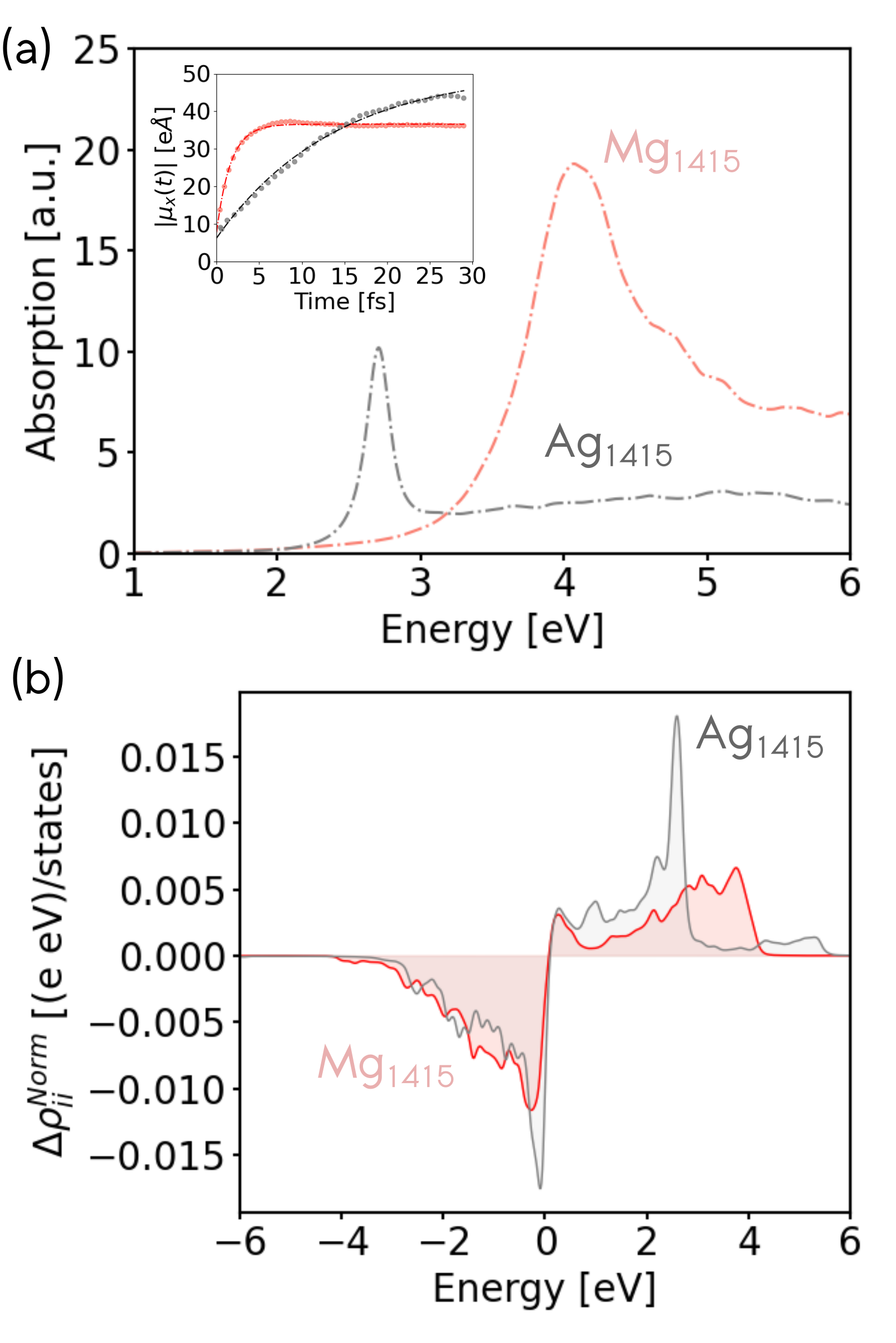}
 \caption{ Optical absorption spectra (top panel) and hot-carriers distribution (bottom panel) for a Ag (black line) and Mg (red line) nanocluster with 1415 metallic atoms. The MO population change for both systems was obtained at 43.6 fs under a constant laser source. Both profiles were normalized for their respective density of states (DOS). Inset: Fitting of the absolute value of the dipole moment signal in  x-direction for a Ag (black line) and Mg  (red line) nanocluster with 1415 metallic atoms when these are illuminated at their respective plasmonic frequencies under constant laser source.}
 \label{fgr:standard}
\end{figure}

We now briefly identify the similarities and differences in optical absorption response between metallic Mg nanoclusters and more conventional plasmonic metals such as Ag.\cite{2016_douglas,2019_douglas,2017_bonafe} The optical and electronic properties of icosahedral Mg and Ag nanoclusters with 1415 atoms are compared in Figure~\ref{fgr:standard}. The optical absorption spectra for both particles are shown in Figure~\ref{fgr:standard}a. The optical response in both metals is characterized by a single clearly defined dipole \added{LSPR} band \added{located at 2.71 and 4.07 eV for Ag and Mg, respectively}. The main spectroscopic features of the Ag nanocluster are largely localised in the visible regime (2-3~eV), whereas for Mg they are in the UV region (3-5~eV). Another key difference is the line width associated with the \added{LSPR} band, which is significantly narrower in Ag than in Mg, associated with a longer lifetime of the \added{LSPR} in the case of Ag (\textit{vide infra}). The absorption is  much more intense for Mg than for Ag. With a strong plasmonic response in the UV range, Mg nanoclusters could potentially complement conventional plasmonic materials, which typically absorb in the visible or near UV range.\cite{2011_hartland,2006_elsayed,2003_elsayed,2000_elsayed,2020_ringe,2018_ringe}

In order to further characterize the plasmonic behaviour of Mg nanoclusters, we report the homogeneous linewidth \added{($\Gamma$)} and lifetime (T$_2$) associated with its plasmonic excitation. The homogeneous linewidth can be calculated by fitting the time-dependent dipole moment response ($\mu(t)$) when the nanostructure is illuminated with a sinusoidal time-dependent electric field as an external perturbation.~\cite{2017_douglas,2016_douglas} 
The absolute value of the dipole moment and its respective fitting curve are shown in the top part of Figure~\ref{fgr:standard} as an inset. The raw dipole moment response for both metals is shown in Supporting Figure S5. The homogeneous linewidth and lifetime for Ag and Mg nanoclusters computed in this way are $\Gamma_{Ag}$ =144.17 meV (9.13 fs) and $\Gamma_{Mg}$ = 1227.61 meV (1.07 fs), respectively. The homogeneous linewidth  computed for the Mg nanoclusters with 1415 metallic atoms is several times larger than that for the Ag nanocluster. This suggests that plasmonic dephasing occurs much more efficiently in nanostructured Mg than in Ag. \added{An important quantity that can be associated with the homogeneous linewidth ($\Gamma$) is the quality factor (Q) which is defined as Q = $E_{LSPR}/\Gamma$.\cite{2002_sonnichse,2015_schatz} This quantity accounts for the extent of the local field enhancement ($\sim Q^4$) and can be used as a metric to judge the performance of a material for plasmonic applications such as  surface-enhanced Raman spectroscopy (SERS) and near-field fluorescence quenching.\cite{2002_sonnichse,2015_schatz} The computed values for Ag and Mg nanoclusters with 1415 atoms are Q$_{Ag}$ = 18.80  and Q$_{Mg}$=3.32. These values indicate that the Ag nanocluster can potentially produce higher local field enhancement than Mg. On the other hand, the Mg nanoclusters show a LSPR in the UV region that efficiently produces high energy hot electrons where Ag or Au are plasmonically inactive. Therefore, specifically in the context of hot electron chemistry, Mg  may represent a suitable alternative to conventional plasmonic materials.}

Finally, we compare the ability to generate energetic EHPs for Mg and Ag nanoclusters in Figure~\ref{fgr:standard}b. Ag posseses a highly localised DOS below the Fermi level associated with the d band, whilst its DOS above the Fermi level is close to constant. Mg on the other hand shows a DOS that increases monotonically with the square root of energy as is expected for a metal with almost free electrons (see Supporting Figure S1). For a fair comparison between both metals, the hot-carrier distribution was normalized by their respective DOS computed at the DFTB level ($\Delta\rho^{Norm}_{ii}$). The non-normalized hot-carrier distribution is shown in Supporting Figure S6. Both metals exhibit a similar distribution of hot holes, but differ in the hot electrons that can be generated.  In the case of Mg, the highest concentration of hot electrons is produced in the range of 3-4~eV, whereas Ag produces a high concentration of hot electrons at $\sim$2.5~eV. In summary, the \added{LSPR} band of Mg nanoclusters has a much shorter lifetime than Ag and produces a broader distribution that reaches electron energies of up to 4~eV.

\subsection{\label{sec:hydrogen} Plasmonic enhancement of hydrogen dissociation on Mg nanoclusters}

Metallic Mg is highly reactive and readily forms a surface hydride when exposed to molecular hydrogen (MgH$_2$).~\cite{2006_johansson,2018_sterl,2012_buckley,2008_pozzo} This property has been discussed in the literature in the context of hydrogen storage.\cite{2020_sterl,2016_sterl,2018_ringe,2020_ringe} In the following, we will interpret the predicted optical properties of Mg nanoclusters in the context of plasmonic enhancement of hydrogen dissociative adsorption and associative desorption (hydrogen evolution). In the previous sections, we have shown that \added{LSPR} excitation of Mg nanoclusters produces hot electrons with energies of up to 4~eV. By analysing the electronic structure and nonadiabatic coupling of adsorbed hydrogen on Mg, we will study if these hot electrons can efficiently couple with adsorbate degrees of freedom.

\begin{figure}[h]
 \centering
 \includegraphics[width=3.3in]{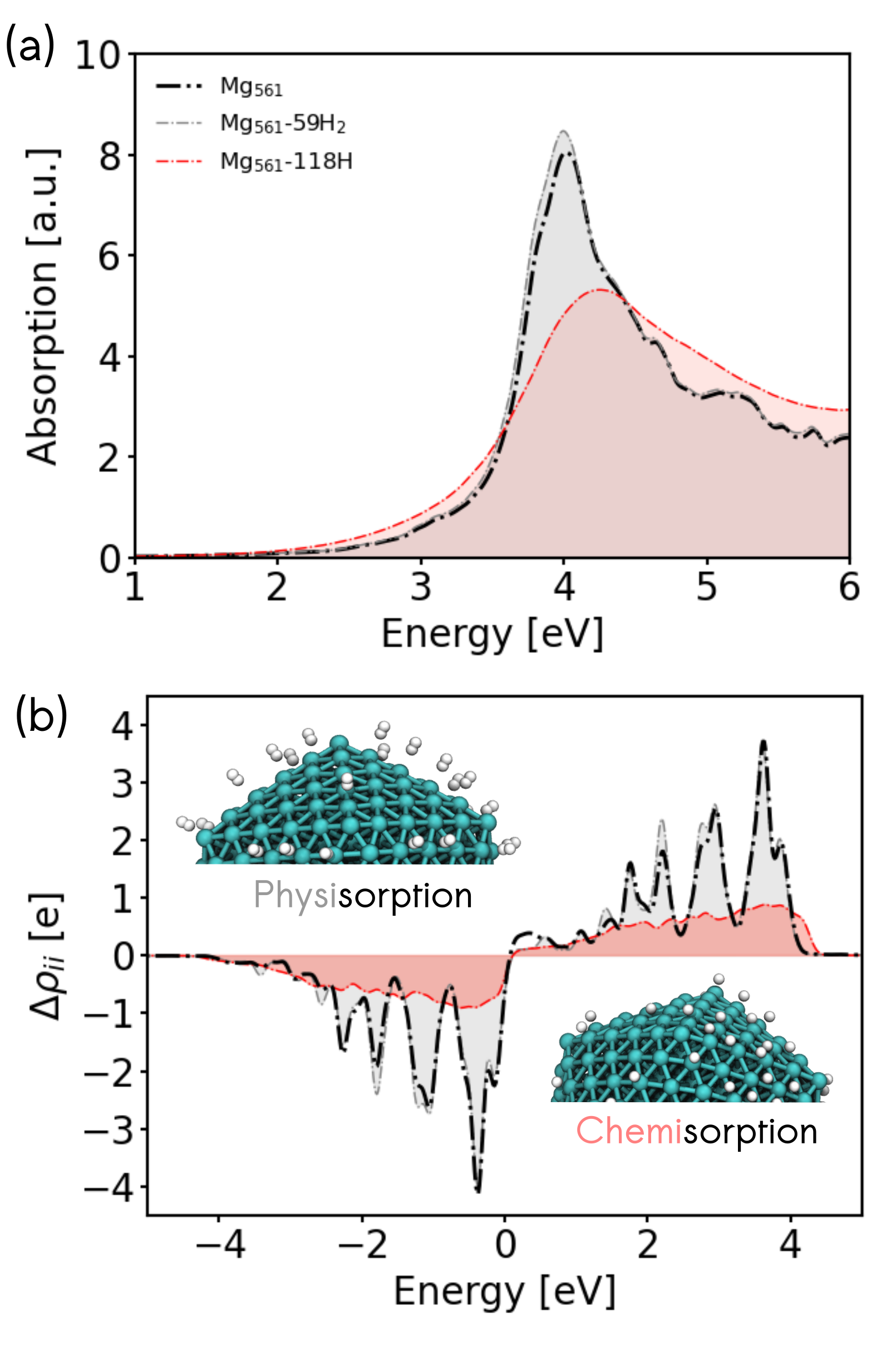}
 \caption{ (a) Optical absorption spectrum and (b) hot-carrier distributions for a naked Mg nanocluster (dotted black line) with 561 metallic atoms, a Mg nanocluster with 59 H$_2$ molecules adsorbed (dotted gray line), and with 118 adsorbed H atoms (dotted red line). The hot-carrier distributions were computed after 96.75 fs of electron dynamics with a constant laser source at the respective plasmonic frequency. }
 \label{fgr:mgh2-hot}
\end{figure}

We first turn our attention to the effect of hydrogen adsorption on the optical properties of Mg nanoclusters. Figure~\ref{fgr:mgh2-hot} shows how hydrogen adsorption changes the optical absorption spectrum and hot carrier distributions of a 561 atom Mg nanocluster. We compare the absorption spectrum of the bare cluster, with the spectrum obtained for clusters covered by physisorbed molecular hydrogen H$_2$ and chemisorbed hydrogen atoms. We choose 59 H$_2$ molecules and 118 H atoms to decorate the nanoparticle at random set of positions that would be filled by the metal atom shell of the next largest icosahedral particle (923 Mg atoms). 
In all cases, stable absorption geometries have been obtained by relaxing the structures of the adsorbate atoms while keeping the previously optimized Mg atoms frozen in their positions. \added{This allows a consistent comparison with the naked Mg nanoclusters by avoiding any surface relaxation effects. The optical and electronic properties associated with a full optimization of Hydrogen covered nanocluster are discussed in the supplemental material in Figure S11. We stress that determining the H adsorption positions on Mg is a challenge.\cite{2015_zhang}  Our approach represents an initial assessment to evaluate the effect of H adsorption on the plasmonic properties of Mg. }


In the case of molecular hydrogen physisorption, we find almost no effect on the \added{LSPR} band or on the EHP distribution. As can be seen in Figure~\ref{fgr:mgh2-hot}b, the hot-electron  distribution for the physisorbed case still leads to  hot-electrons with energies of up to $\sim$4~eV. On the other hand, a very strong CID effect is found in the case of chemisorbed hydrogen atoms on Mg,  which reduces the \added{LSPR} lifetime (increase of the linewidth) and induces a blue-shift of the maximum of the \added{LSPR} band. This optical behavior is opposite to what was found for Ag nanoclusters, where strongly hybridised adsorbates produced a red-shift of the \added{LSPR} band.~\cite{2016_douglas}
Likewise, the chemisorption of hydrogen atoms reduces the number of hot electrons produced (Figure~\ref{fgr:mgh2-hot}b), but reaches slightly higher hot electron energies.

Molecular hydrogen dissociation is a highly activated process on Mg. We have calculated a minimum energy path for the dissociative adsorption reaction of a small Mg nanocluster (55 atoms) and a p(3x3) Mg(0001) periodic surface slab using DFT (See Figure~\ref{fgr:slabs}a and Supporting Figure~S7a). Along this path, we identify three key geometries, the initial physisorbed state (IS), the transition state (TS), and the final chemisorbed state (FS). We find an activation energy of 0.71~eV and 0.89~eV for hydrogen dissociation on the nanocluster and the surface, respectively. This is in good agreement with a measured activation energy of  0.75$\pm$0.15~eV)\cite{2006_johansson} and with other theoretical reports.~\cite{2008_pozzo,2009_pozzo,2013_lei} We further analyse the electronic structure of the 3 key geometries by visualising the projected DOS (pDOS) associated with a single physisorbed  H$_2$ molecule on Mg nanoparticle and on a clean Mg(0001) surface in Supporting Figure S8. In both cases, the IS shows little hybridisation and state coupling. The unoccupied $\sigma^*$ state of $H_2$ shows some level of state splitting in the range of 3 to 6~eV above the Fermi level, which suggests some coupling to metallic states. 
This is in contrast to the FS, where the projected DOS shows that the electronic states of adsorbed hydrogen atoms strongly couple to the metallic states of Mg leading to a homogeneous distribution of H contribution across a wide range of energies. 

For plasmonic enhancement of hydrogen dissociation on Mg nanoclusters, hot electrons need to be able to effectively transfer into the antibonding $\sigma^*$ MO of a physisorbed H$_2$ molecule, which requires sufficient molecule-metal coupling and alignment of the molecular levels with the energetic distribution of hot electrons created by \added{LSPR} decay. Both requirements are reached for Mg nanoclusters. Hot electrons are efficiently created with high energies of 3-4~eV for a wide range of nanocluster sizes, which is an energy region that overlaps with unoccupied adsorbate states. Therefore, our calculations point towards effective plasmonic enhancement of hydrogen dissociation on Mg nanoclusters and surfaces, which is yet to be corroborated in experiments. 

\subsection{\label{sec:HER} Plasmonic enhancement of hydrogen evolution from Mg nanoclusters}

To address the question if plasmonic excitation can also enhance hydrogen evolution, we need to study the reverse reaction, namely the recombination of chemisorbed H atoms into molecular hydrogen. Here, the situation is less clear as the electronic states of chemisorbed hydrogen atoms are heavily hybridised and homogeneously embedded into the DOS of the metal (see Supporting Figure~8). Therefore, there are no clear resonances which promote the reaction. This is a case that is well described by Fermi's golden rule where EHPs will couple with hydrogen atom motion proportional to the magnitude of the DOS and the nonadiabatic coupling strength between molecular motion and electronic states.~\cite{Head-Gordon1992,2016f_maurer} In this limit, nonadiabatic energy dissipation effects during dynamics on surfaces can be described by molecular dynamics with electronic friction (MDEF),~\cite{Head-Gordon1995, 2016_maurer} where we describe the effect of EHPs on molecular motion as frictional and random force governed by an electronic friction tensor. This approach is well justified and common for the study of hydrogen atom dynamics on metals~\cite{Rittmeyer2016, dorenkampHydrogenCollisionsTransition2018} and has also been applied to study light driven diffusion and desorption dynamics in combination with TTM.~\cite{Loncaric2016,Juaristi2017} Using first order time-dependent perturbation theory on Kohn-Sham DFT, we can calculate the relaxation rates and vibrational lifetimes associated with different directions of atomic and molecular motion along the minimum energy path.~\cite{2017_maurer}

\begin{figure}[h]
 \centering
 \includegraphics[width=3.3in]{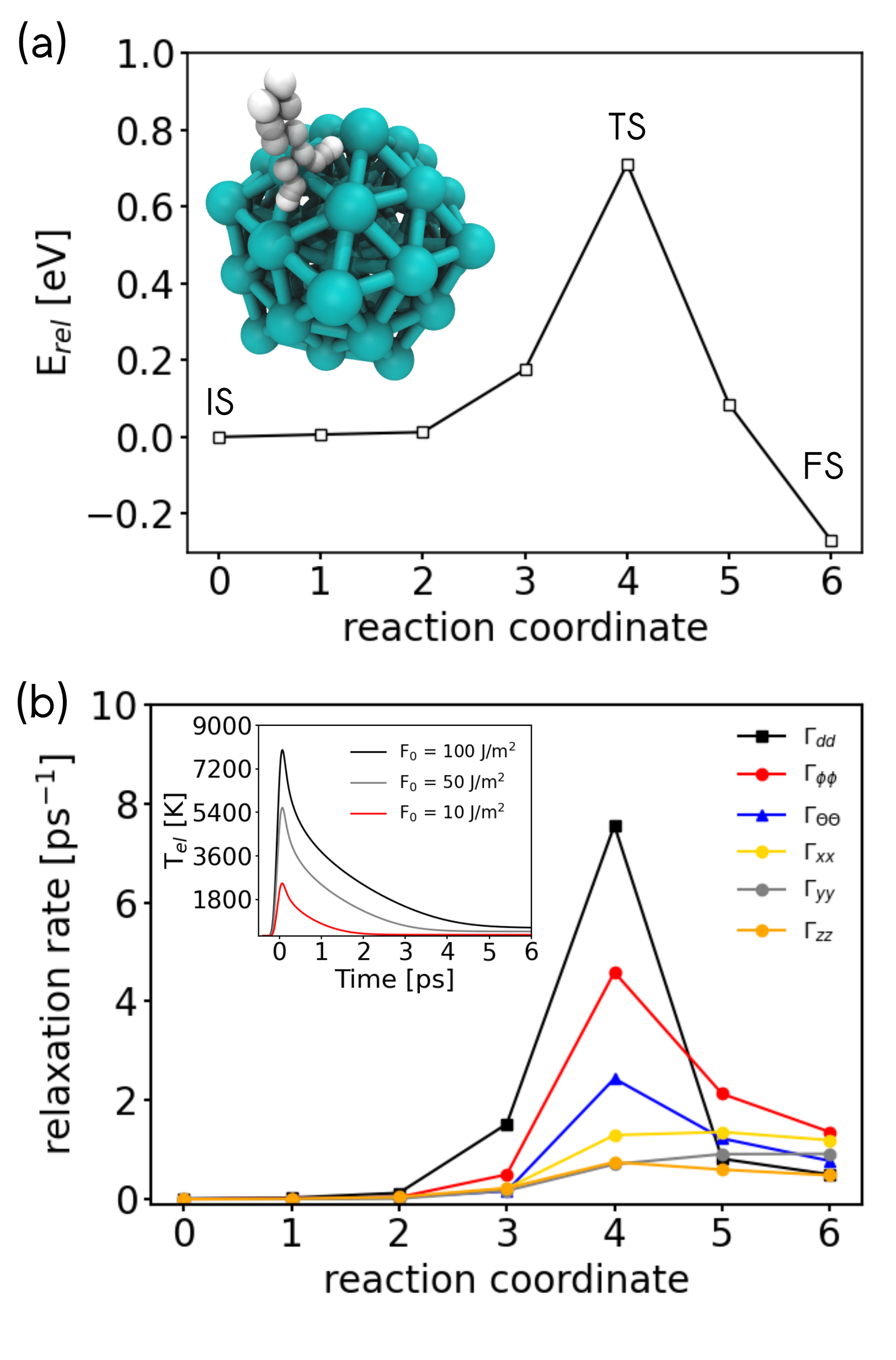}
 \caption{ (a) Minimum energy path (MEP) for the molecular hydrogen dissociation reaction on a Mg nanocluster with 55 metallic atoms and the (b) vibrational relaxation rates of molecular adsorbate motion due to hot electrons for geometries along this reactive path. Rates are given along internal coordinates of the molecule defined in Ref.~\cite{2021_box} Inset: Electronic temperature computed for Mg by means of TTM at three different laser fluences.}
 \label{fgr:slabs}
\end{figure}

In Figure~\ref{fgr:slabs}b and Supporting Figure S7b, we report the nonadiabatic relaxation rates of hydrogen motion due to coupling with EHPs along the minimum energy path of H$_2$ dissociation. Similar to the previously reported case of H$_2$ dissociation on Ag(111), we find the strongest coupling to EHPs for the intramolecular stretch motion of H$_2$ ($\Gamma_{dd}$) at the TS,~\cite{2017_maurer} however the relaxation rate on the Mg nanocluster and the Mg(0001) surface is significantly higher reaching about 8~ps$^{-1}$, which corresponds to a vibrational lifetime of the internal stretch of about 100~fs. In the FS, nonadiabatic relaxation rates are comparable between Ag(111) and Mg(0001) with the biggest components of the relaxation rate tensor corresponding to about 2~ps$^{-1}$ (or a lifetime of 0.5~ps). 

We can use this number for an approximate calculation of how much energy \added{LSPR} excitation can transfer into an adsorbed hydrogen atom. We do this by performing Langevin dynamics within MDEF where we neglect the underlying potential energy surface and take the friction coefficient that describes the coupling between hot electrons and adsorbate as constant. We couple these equations with the simulated TTM temperature profiles shown in the inset of Figure 5 to mimic how the electronic temperature changes as a function of time. By propagating the relevant equation in time, we can calculate the kinetic energy that will be transferred from the EHPs into the H atoms. (for details see supporting section S9).~\cite{Leimkuhler2013} As shown in the inset of Figure~\ref{fgr:slabs}b, depending on the laser fluence, electronic temperatures of 2000-8000~K can be generated (see supporting section S8 for details). According to the relaxation rates we find for the FS in Figure 5, we select a friction coefficient that corresponds to a relaxation rate of 2~ps$^{-1}$. We find that a laser pulse with a \added{raw laser} fluence of 10~J/m$^2$ (100~J/m$^2$) leads to peak electronic temperatures of 2500~K (8000~K) and an increase of hydrogen kinetic energy by 0.17~eV (0.55~eV) over 250~fs. The DFT-PBE predicted energy required to overcome the barrier for desorption is 0.71~eV and 0.89~eV on the 55 atom nanocluster and the Mg(0001) surface slab, respectively.  These results suggest that significant H$_2$ desorption should be observed for fluences between 50-100~J/m$^2$ as two H atoms will be able to gain sufficient energy to overcome the barrier for hydrogen evolution. Fluences in this range can be achieved by laser irradiation and have been reported previously in ultrafast laser driven photodesorption studies.\cite{r2_Ertl,r2_funk,r2_saalf} We note that this simple calculation neglects the underlying energy landscape and other effects, such as the dependence of nonadiabatic coupling rates $\gamma$ on the coordinates and the electronic temperature.~\cite{2016f_maurer,Maurer2019Faraday}

\section{Conclusion}
Earth-abundant materials are becoming increasingly important as catalysts as they are more cost effective than traditional catalyst materials. In the context of plasmonic catalysis, Mg nanoclusters may provide interesting complementary properties to conventional plasmonic materials such as gold and silver. In this work, we study the plasmonic properties of Mg nanoclusters using electronic structure and TD-DFTB calculations, with a particular focus on how the plasmonic behaviour could be used to catalyze hydrogen absorption and release in the context of hydrogen storage. 

Electron dynamics simulations show that Mg nanoclusters are able to produce hot-carrier distributions with hot electrons at energies of $\sim$ 4 eV. Therefore Mg nanoclusters produce higher concentrations of high energy hot electrons than silver nanoclusters of the same size. We have studied the optical absorption properties and the hot-carrier production on pristine and hydrogen-covered Mg nanoclusters of various size. By analysis of the electronic structure of physisorbed molecular hydrogen and chemisorbed atomic hydrogen on Mg, we conclude that it is highly likely that hydrogen dissociation can be selectively promoted by plasmonic excitation of Mg nanoclusters. On the other hand, hydrogen evolution from Mg nanoclusters is expected to be less sensitive to plasmonic excitation but generally effective in the presence of high temperature thermalised hot electrons \added{at least for an early H adsorption stage where the plasmonic properties and metallic character are still held}. This scenario can occur during constant illumination or local surface heating. As the dissociative adsorption and hydrogen evolution reactions couple to hot electrons differently, we speculate that it may be possible to realise experimental strategies where one or the other are more preferentially promoted. Our results confirm that nanostructured metallic Mg is a promising plasmonic material for applications in photocatalysis and hydrogen storage.


\section{\label{sec:methods}Computational Methods}

\subsection{Electronic structure of Mg nanoclusters}

The ground-state electronic structure of Mg nanoclusters has been computed using a self-consistent-charge density-functional-tight-binding (SCC-DFTB) approach.\cite{1998_elstner,2014_elstner,2016_elstner} This method is based on a second order expansion of the Kohn-Sham DFT total energy around a non-interacting reference density built from a  superposition of atomic densities.\cite{1998_elstner, 2014_elstner,2016_elstner} The SCC-DFTB method has been successfully employed to describe the electronic structure and quantum properties of large organic, inorganic and  biological systems in the past.\cite{2016_elstner,2014ap_elstner,2020_rapacioli} This semi-empirical method can overcome inherent computational limitations associated with DFT calculations.\cite{2014ap_elstner,2016_elstner,2018_aikens,2020_rapacioli} We employ the DFTB+ package\cite{2007_aradi,2020_dftb+} to describe the ground-state and time-dependent electronic properties of Mg nanoclusters. The 3ob-3-1 DFTB\cite{2015_3ob} parameter set is used and the reliability in the predicting band structure and DOS for bulk Mg has been verified against DFT calculations using the FHI-Aims\cite{2009_fhi-aims} quantum chemistry package and the Perdew, Burke, and Ernzerhof (PBE) functional\cite{pbe} (see validation  section in ESI and Supporting Figure S1). 
This validation study was also extended to compute the density of states (DOS) associated with very small Mg nanoclusters containing 13, 55, 147, 309  atoms (see Supporting Figure S2).
\subsection{Electron dynamics}

The 2020 release version of DFTB+ has been used\cite{2020_dftb+, 2020_bonafe} to simulate the electron dynamics associated with the plasmonic excitation and thus to obtain the optical absorption spectra and other related dynamics properties such as the hot-carrier distribution and the plasmonic lifetime.  This is made possible with a new implementation to describe the electron dynamics based on time-dependent SCC-DFTB (TD-SCC-DFTB).\cite{2020_bonafe,2020_dftb+}
\added{Within this theoretical framework, the electron dynamics is driven under the influence of an external time-varying potential $(V_{ext} (t))$ or electric field ($E(t)$)}

\added{\begin{equation}
    H(t)=H_{GS}+V_{ext}(t)= H_{GS}-E(t)\mu
 \end{equation}}

\added{Where $H_{GS}$ is the ground state Hamiltonian and $\mu$ is the dipole moment operator.\cite{2020_dftb+,2020_bonafe}} Two different kinds of \added{external electric field} have been used to drive the electron dynamics. For the optical absorption spectra, a Dirac $\delta$ pulse was used as external perturbation ($E(t)=E_0\delta (t-t_{0})$). \added{The optical absorption spectrum is proportional to the imaginary part of frequency-dependent dynamic polarizability ($\alpha(\omega)$) which can be obtained from a Fourier transform of the time-dependent dipole moment, within the linear response regime (small electric fields). To determine the plasmonic resonance (LSPR peak) frequency, we select the main spectroscopic resonance by inspection. The LSPR band is often the most intense peak within the computed optical absorption spectrum.} To explore the photophysics associated with plasmonic excitation, a sinusoidal time-dependent electric field \added{($E(t)=f(t)E_0 sin (\omega_{LSPR}^{Mg} t)$)}  in tune with the plasmonic frequency was selected as external perturbation to drive the electronic dynamics and to calculate different emerging dynamical properties such as the plasmonic excitation lifetime and the hot-carrier distribution.
\added{The $f(t)$ function is used to modulate the external electric field for the second type of external perturbation. Two different shapes have been considered, a continuous laser source with $f(t)=1$ and a  pulsed laser source with $f(t)= sin^{2}( \pi(t-t_0)/ \tau)$ if $t_0 < t < t+\tau$  and $f(t)=0$ otherwise.\cite{2020_dftb+} }All nuclear positions have been frozen at their respective optimized geometries during the electron dynamics simulations. 

\subsection{CI-NEB and electronic friction calculations}
The non-adiabatic effects during H$_2$ dissociation on metallic Mg have been characterized by computing the electronic friction tensor elements ($\Lambda_{ij}$) for geometries along the MEP for two different systems, a periodic Mg(0001) slab and a small Mg nanocluster.\cite{2016f_maurer,2017_maurer} To compute the electronic friction tensor, the current implementation within FHI-aims has been used.\cite{2009_fhi-aims,2016f_maurer} This implementation calculates nonadiabatic coupling matrix elements to calculate the relaxation rates on adsorbate motion due to coupling with hot electrons in the metal. The MEP was obtained by using the  climbing-image nudged elastic band (CI-NEB) method implemented in ASE\cite{2017_ase} with FHI-aims. The MEP path contains 7 and 9 images for the Mg nanocluster and periodic Mg(0001) slab, respectively. The set of images was  optimized with a maximum force threshold of 0.05 eV/$\AA$. For the Mg nanocluster case, a small iscohedral Mg particle containing 55 metallic atoms was considered. For the periodic Mg(0001) slab a $3\times3$ surface unit cell  with 4 atomic layers (36 atoms) was employed with the bottom two atomic layers frozen in order to retain bulk properties. A Mg lattice constant for PBE previously published (a=3.19 c=5.12)\cite{2007_lattice-constant} was employed for the periodic system. The calculated lattice constants are in good agreement with the experimental ones (a =3.21$\mathrm{\AA}$  and c=5.21 $\mathrm{\AA}$).\cite{1957_lattice-exp} To compute the MEP and the electron friction components, a Monkhorst-Pack k-point mesh of $12\times12\times1$  was chosen for the periodic systems. This particular k-point grid produced converged results for the relaxation rates associated with the electronic friction elements (see Supporting Figures S7c and S7d). We use a broadening of the electronic states of 0.6 eV when calculating relaxation rates. The details of this procedure are explained in Ref.~\cite{2016f_maurer} In both studied systems, PBE and "tight" basis set option were chosen as suitable choices to carry out these calculations. 


\section*{Acknowledgements}
O.A.D-G and R.J.M acknowledge funding from the UKRI Future Leaders Fellowship programme (MR/S016023/1). C.L.B. is supported by an EPSRC-funded Ph.D. studentship. High performance computing resources were provided via the Scientific Computing Research Technology Platform of the University of Warwick, the EPSRC-funded Materials Chemistry Consortium for the ARCHER UK National Supercomputing Service (EP/R029431/1), and the EPSRC-funded HPC Midlands+ computing centre (EP/P020232/1). The authors want to thank James Gardner \added{and Franco P. Bonafé} for fruitful discussions.

\section*{Author contributions}

R.J.M. supervised the project. R.J.M. and O.A.D-G. conceived the project. O.A.D-G. carried out TD-DFTB and DFT calculations. C. L. B. and R.J.M. carried out electronic friction calculations. O.A.D-G., C. L. B. and R.J.M. discussed and interpreted the data and wrote the manuscript.

\section*{Competing Interests}
The authors declare no competing interests.

\section*{Data Availability}
The input and output files of all electronic structure calculations have been uploaded to the NOMAD repository and are available as a data set at doi: \href{ https://doi.org/10.17172/NOMAD/2021.03.29-1}{10.17172/NOMAD/2021.03.29-1}. The data shown in all figures has been uploaded to figshare at doi: \href{https://doi.org/10.6084/m9.figshare.14339687}{10.6084/m9.figshare.14339687}.


\providecommand*{\mcitethebibliography}{\thebibliography}
\csname @ifundefined\endcsname{endmcitethebibliography}
{\let\endmcitethebibliography\endthebibliography}{}

\end{document}


\title{Supplemental Material for \\
"Plasmonic enhancement of molecular hydrogen dissociation on metallic magnesium nanoclusters"}

\author{Oscar A. Douglas-Gallardo}
\affiliation{Department of Chemistry, University of Warwick, Gibbet Hill Road, Coventry, CV4 7AL, UK.}
\author{Connor L. Box}
\affiliation{Department of Chemistry, University of Warwick, Gibbet Hill Road, Coventry, CV4 7AL, UK.}
\author{Reinhard J. Maurer}
\email{r.maurer@warwick.ac.uk}
\affiliation{Department of Chemistry, University of Warwick, Gibbet Hill Road, Coventry, CV4 7AL, UK.}

\date{}





{
\let\clearpage\relax
\maketitle
}



\tableofcontents


\section{Electronic structure benchmark calculations}

The ground-state and time-dependent electronic properties associated with Mg nanoclusters have been computed with the SCC-DFTB method\cite{2007_aradi} and the 3ob-3-1\cite{2015_3ob} parameter set. These DFTB parameters were originally parametrised for organic and biological systems, thus its direct transferability to other molecular and metallic systems needs verification. In this context, a set of DFT calculations  have been carried out to compare them with the DFTB results, see Figure S1. This validation study shows that DFTB is able to generate reliable results reproducing the general electronic structure features computed at DFT-PBE\cite{pbe} level. A  small error in the lattice constants at the DFTB level is detected for the optimized unit cell ($\Delta c$=12.9 and $\Delta a$ 5.3$\%$).

\begin{figure}[h!]
 \centering
 \includegraphics[width=15.0cm]{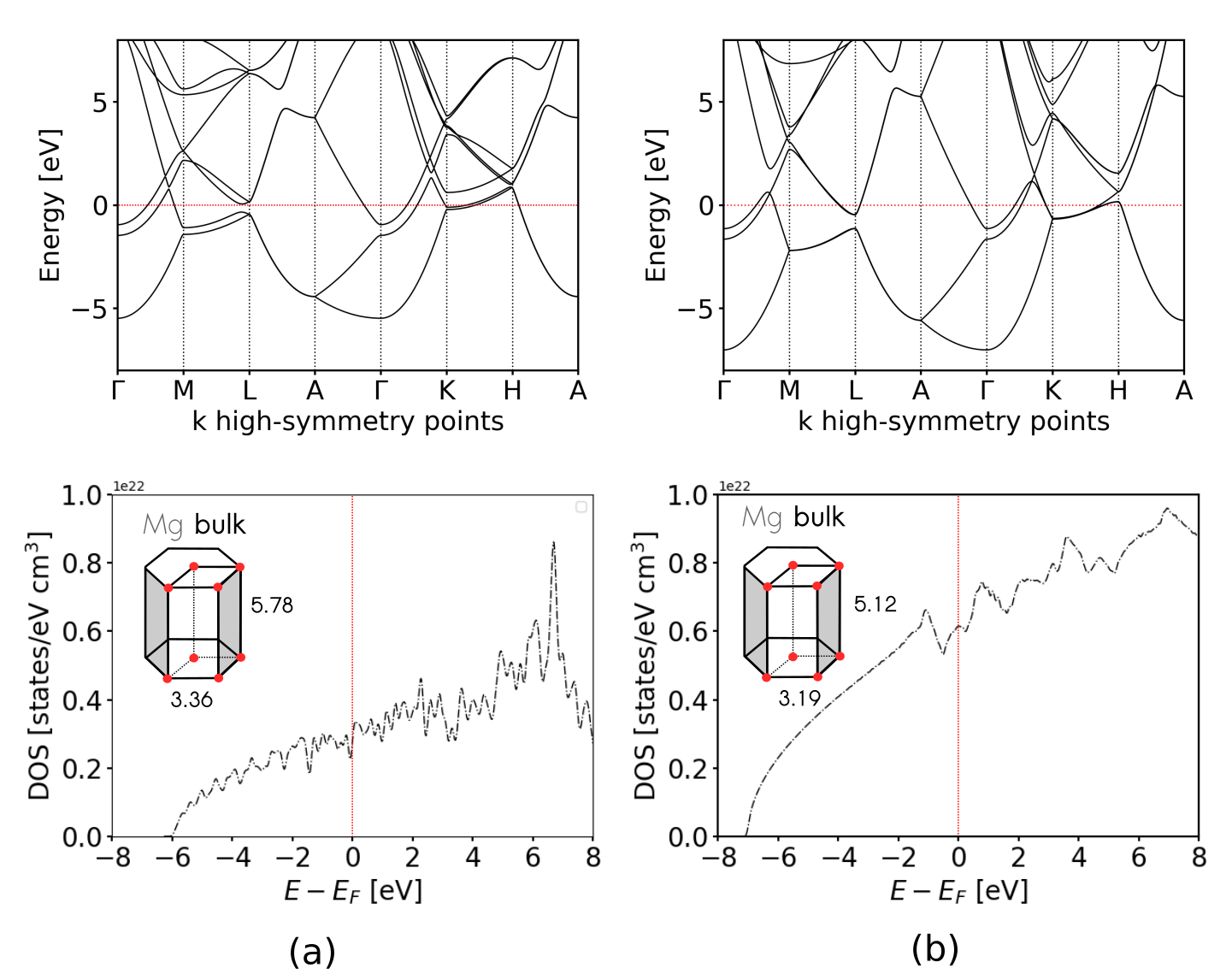}
 \caption{ Band structure and  DOS for Mg bulk computed  at (a) DFTB (3ob-3-1) (b) DFT-PBE level of theory. The dotted red lines indicate the Fermi energy in all panels. Inset: Unit cell parameters in \AA ngstrom obtained after unit cell optimization are shown for Mg hcp crystal structure at DFTB  and DFT level. To compute the band structure, a k-grid 16x16x16 was selected. In the case of the DOS a denser k-grid was used ($96\times96\times96$) with a broadening width of 0.05 eV (the same set up was used for DFTB).}
 \label{fgr:dos-band-validation}
\end{figure}

We have calculated the Mg band structure and Mg density of state (DOS) and compared them with DFTB results.  Both properties emerge from the underlying electronic structure and can be used as suitable metric to assess the quality of these parameters. To compute the Mg band structure, a collection of high-symmetry k points on hexagonal closest packed (hcp) structure\cite{2010_kpoints,1962_brian,2007_chulkov} were selected ($\Gamma$, M, L, A, K and H).\cite{2010_kpoints} The top panel of Figure S1 shows a very good agreement between the band structure of bulk Mg predicted by DFT and DFTB  and both  compare fairly well with other reports.\cite{1962_brian,2007_chulkov} The band structure at DFT level is characterized by a set of flat bands very close to the Fermi energy (dotted red line) which are quite well described at DFTB level. The energy of the $\Gamma$ and A points are lower ($<$ -6 eV) at DFT level than the computed ones at DFTB.
The band structure at DFT level was computed with the FHI-aims electronic structure package\cite{2009_fhi-aims} with a k-grid of $16\times16\times16$, a 'tight' basis set and a lattice constant for metal Mg at PBE level of a=3.19 $\mathrm{\AA}$ and c=5.12 $\mathrm{\AA}$.\cite{2007_lattice-constant} These values are in good agreement with the experimental values(a =3.21$\mathrm{\AA}$  and c=5.21 $\mathrm{\AA}$).\cite{1957_lattice-exp} The unit cell employed in the DFTB calculations was first optimized (see inset in bottom panel of Figure S1a for optimized values) using 3ob-3-1 DFTB parameter set. SCF accuracy thresholds employed for the electron density, eigenvalue energies, total energy and force threshold were 1e$^{-6}$ $\mathrm{e/a_0^3}$, 1e$^{-3}$ eV, 1e$^{-6}$ eV and 1e$^{-4}$ $\mathrm{eV/\AA{}}$ respectively.
This good agreement between DFT and DFTB found in the band structure is also extended to the DOS as can be seen in the bottom panels of Figure S1, both show similar profiles and very similar behavior to the free-electron model, which can be assumed for metallic Mg. The DOS was computed by using a denser k-grid ($96\times96\times96$) considering 4000 points (each 0.01eV) and with a broadening of 0.05 eV. Both DOS have been normalized by their respective unit cell volume in order to make a fair comparison. 


\section{Density of states of Mg nanoclusters}
As part of our validation study, the DOS  for a set of small iscosahedral Mg nanoclusters was also computed at DFTB and DFT level, shown in Figure S2. The agreement between DFT and DFTB is generally good, with the general shape and energy range of the DOS captured for a range of particle sizes.

\begin{figure}
 \centering
 \includegraphics[width=11cm]{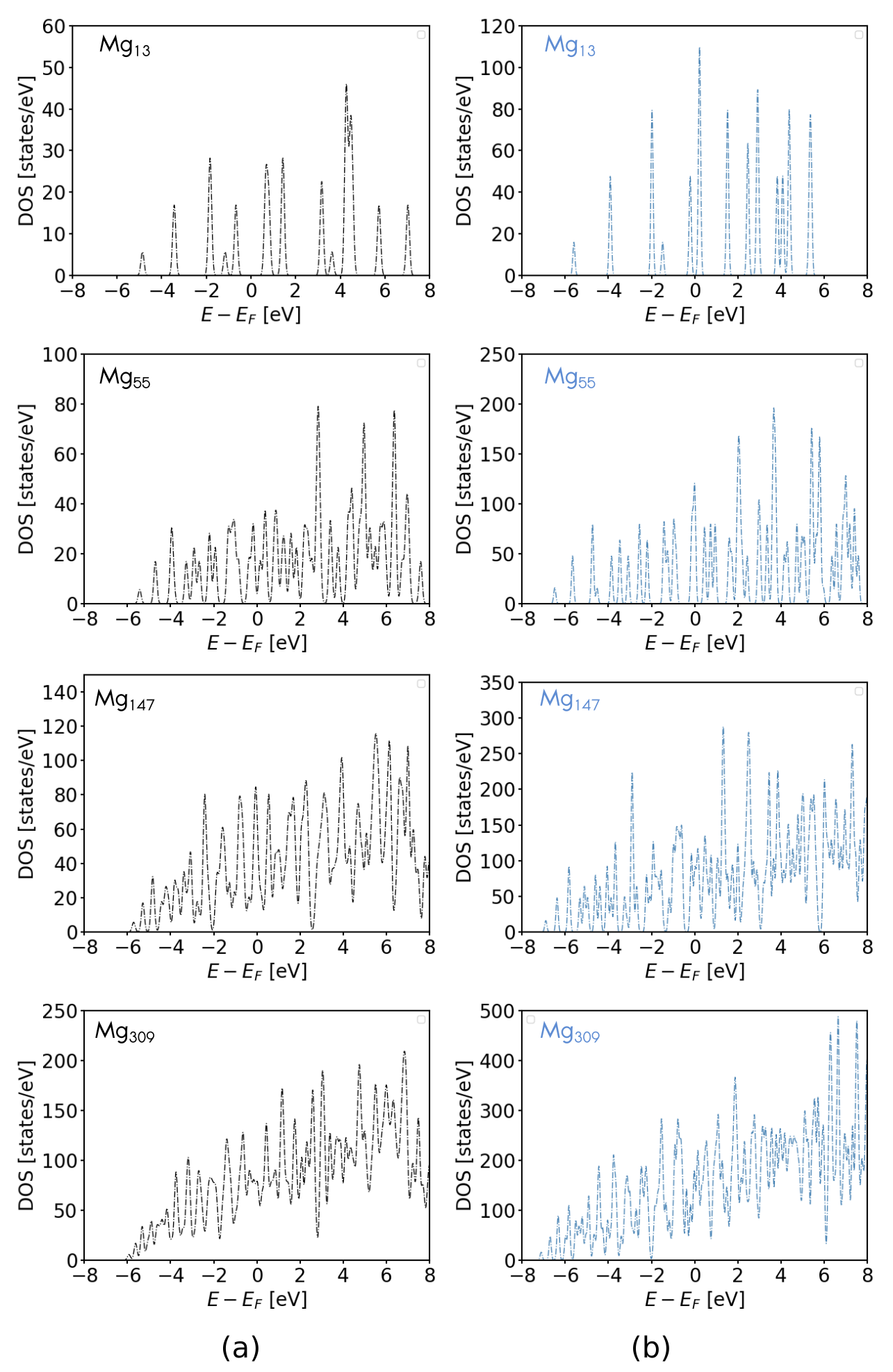}
 \caption{DOS for a set of very small Mg nanoclusters containing  13, 55, 147 and 309 metallic atoms computed with (a) DFTB and (b) DFT. The same set up as described in Figure S1 was used to compute the DOS for these systems (but without a k-grid)}
 \label{fgr:zone-all}
\end{figure}

\newpage
\section{Non-equilibrium electronic distributions}

In Figure S3, we fit the non-equilibrium electron distributions calculated with TD-DFTB with a quasi-logarithmic function $\Phi$(E,t) \cite{2013_rethfeld,2017_rethfeld}, for different time steps of the electronic dynamics for both types of external electric fields.  The $\Phi$(E,t) is defined as follows:
\begin{equation}
   \Phi[E,t]=\Phi[\rho(E,t)]=-\mathrm{ln}[\frac{2}{\rho(E,t)}-1]
\end{equation}
where $\rho(E,t)$ is the electronic population distribution at any time t and $\rho(E,0)$=$f(E,0)$ is the Fermi-Dirac distribution at 300 K.
\begin{figure}[!h]
 \centering
 \includegraphics[width=12cm]{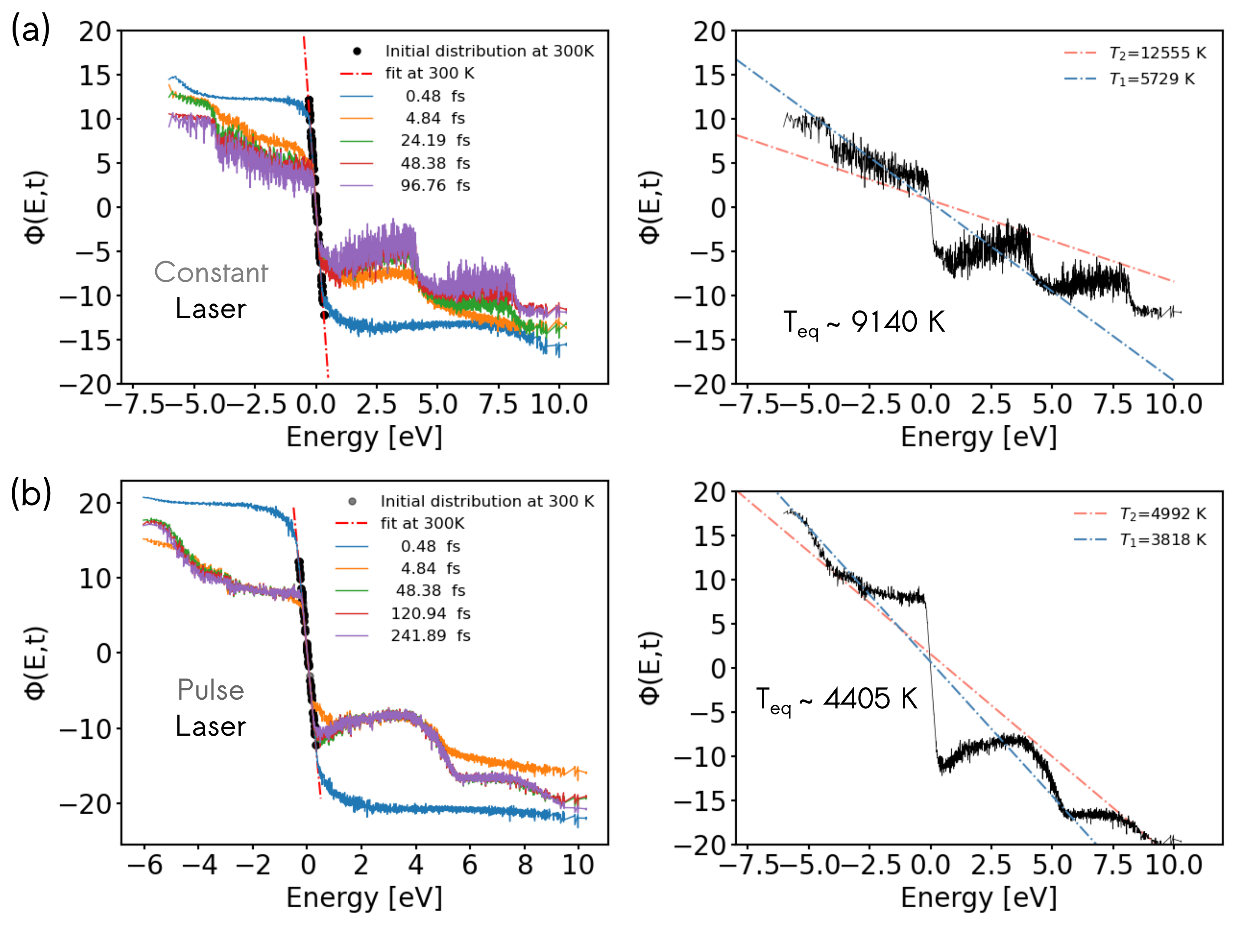}
 \caption{ Quasi-logarithmic representation of non-equilibrium distribution computed at different steps of the electronic dynamics for both (a) constant and (b) pulsed laser sources for the 1415 atom Mg nanocluster. The  right panels further show a rough estimation of the electronic temperature by performing a linear fit to a Fermi-Dirac distribution. The blue dash-dotted lines correspond to fits with minimal residuals. The red dash-dotted fits correspond to fits on the extrem step edges of the distribution. This is to better estimate the variance of electronic temperatures in this approximate approach that stems from the fact that the system is not indeed in electronic equilibrium.}
 \label{fgr:zone-all}
\end{figure}

For both external electric fields the initial distribution at 300 K is shown with black dots. This initial distribution can be very well described by a Fermi-Dirac distribution function at 300 K (dotted red lines) in this quasi-logarithmic representation. Also, a rough electronic temperature estimation was calculated for both laser sources by computing the slope associated with the external edge of the stepwise structure (see right panels) at 96.76 fs and 241.89 fs, respectively. The obtained final electronic temperatures associated with the equilibrium state (after electron-electron thermalization) turned out to be $\sim$ 9142 $\pm$ 3413~K  and $\sim$ 4405 $\pm$ 587~K for constant and pulsed laser sources, respectively. 

\section{Hot-carrier distribution for different Mg nanocluster sizes}

The hot-carrier distribution generation process was also computed for Mg nanoclusters of different sizes. Figure S4 shows the hot-carrier distribution computed after 96.76 fs when a constant laser source is used as external electric field with a  field intensity of $E_0 = 0.02~V~\AA^{-1}$.

\begin{figure}[!h]
 \centering
 \includegraphics[width=16cm]{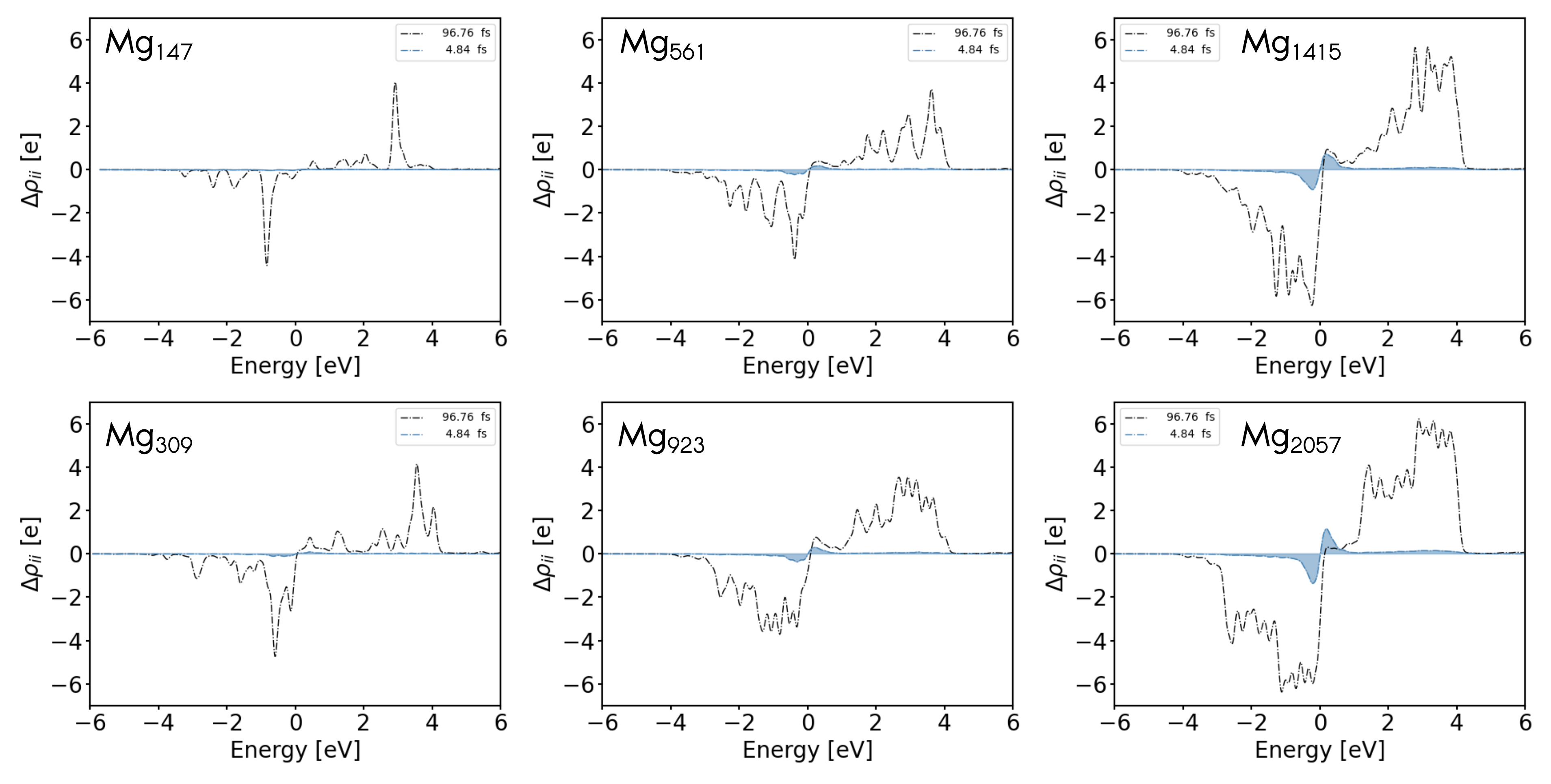}
 \caption{ Evolution of molecular orbitals (MO) population $\Delta\rho_{ii}$ for different particle sizes at \added{4.84 and} 96.76
fs of electronic dynamic by using a constant laser source with an electric field intensity at  0.02 V $\AA^{-1}$. }
 \label{fgr:zone-all}
\end{figure}
\newpage

\section{Comparison of Mg and Ag plasmonic response; dipole moment}
The dipole moment signal was used to extract the linewidth and lifetime associated with plasmonic excitation for both plasmonic metals. Figure S5 shows the dipole moment obtained from our electronic dynamics simulation under constant laser source.

\begin{figure}[!h]
 \centering
 \includegraphics[width=14cm]{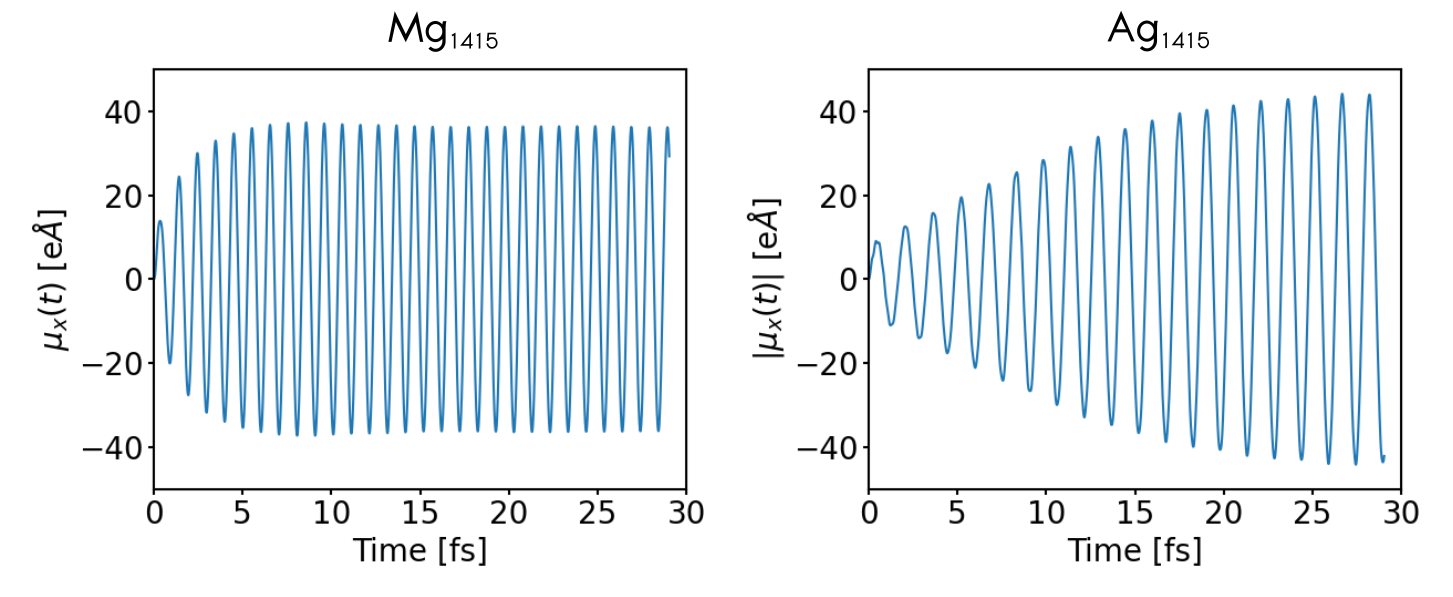}
 \caption{Dipole moment (x-component) obtained from electronic dynamics driven by a constant laser source with a frequency in tune with the plasmonic excitation for a Mg (\added{$\omega_{LSPR}$}= 4.069 eV) and Ag~(\added{$\omega_{LSPR}$}= 2.7098 eV) nanocluster 
with 1415 metallic atoms. In both cases an electric field intensity of $E_0$=0.02~V~$\AA^{-1}$ was used.}
 \label{fgr:zone-all}
\end{figure}

\section{Comparison of Mg and Ag plasmonic response; hot-carriers}

Also, the hot-carriers energetic landscape produced for a Mg nanocluster was computed and  compared with the hot-carrier profile generated by a similar Ag nanocluster. Figure S6 shows the raw  hot-carrier energetic landscape (without normalization by DOS as done in the  main manuscript) obtained from our electronic dynamic simulation at 43.54 fs under constant laser source for both metals.

\begin{figure}[!h]
 \centering
 \includegraphics[width=8cm]{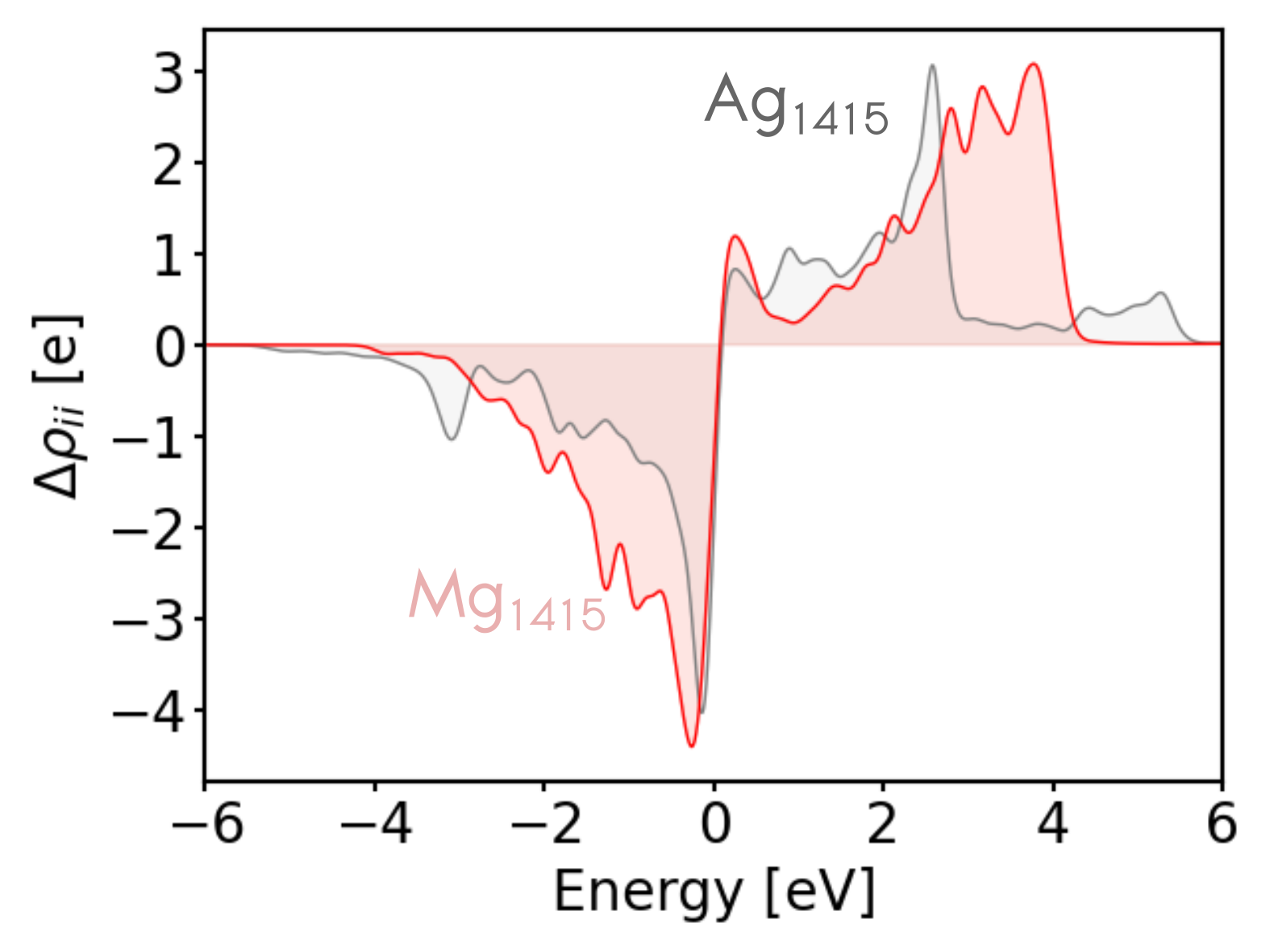}
 \caption{The hot-carrier distribution computed for a Ag and Mg
nanoclusters with 1415 metallic atoms at 43.54 fs. In both cases a constant laser source with an electric field intensity of $E_0$=0.02 V $\mathrm{\AA}^{-1}$ was used.}
 \label{fgr:zone-all}
\end{figure}

\newpage
\section{Minimum energy path and nonadiabatic relaxation rates for a periodic Mg(0001) slab}

The hydrogen dissociation reaction was also explored on a periodic Mg(0001) slab. For this system the minimum energy path (MEP) was computed by using the climbing-image nudged elastic band (CI-NEB) method implemented in ASE and FHI-aims.\cite{CI-NEB,2017_ase} The reaction and activation energies computed at PBE level were -0.02 eV and 0.89 eV, respectively. Likewise, in order to characterize the non-adiabatic effects associated with this chemical reaction, the relaxation rates due to electronic friction were also computed along this reactive path. For each geometry we use first order time-dependent perturbation theory on the Kohn-Sham DFT wavefunctions to calculate the rate of energy loss for each coordinate due to coupling with hot electrons.~\cite{2016f_maurer} We use a broadening of the electronic states of 0.6 eV, a Fermi factor corresponding to an electronic temperature of 300 K and a 'tight' basis set. SCF accuracy thresholds employed for the electron density, eigenvalue energies and total energy were 1e$^{-6}$ $\mathrm{e/a_0^3}$, 1e$^{-3}$ eV and 1e$^{-6}$ eV respectively. The details of this procedure are explained in Ref.~\onlinecite{2016f_maurer} The Figure S8a and S8b show the obtained MEP and the resulting electronic friction relaxation rates for this system. Figures S8c and S8d show the convergence tests for different k-grids at two special points, the TS and FS. This test shows that k-grid $12\times12\times1$ is a robust value to compute the electron friction elements for this periodic system.

\begin{figure}[!h]
 \centering
 \includegraphics[width=14cm]{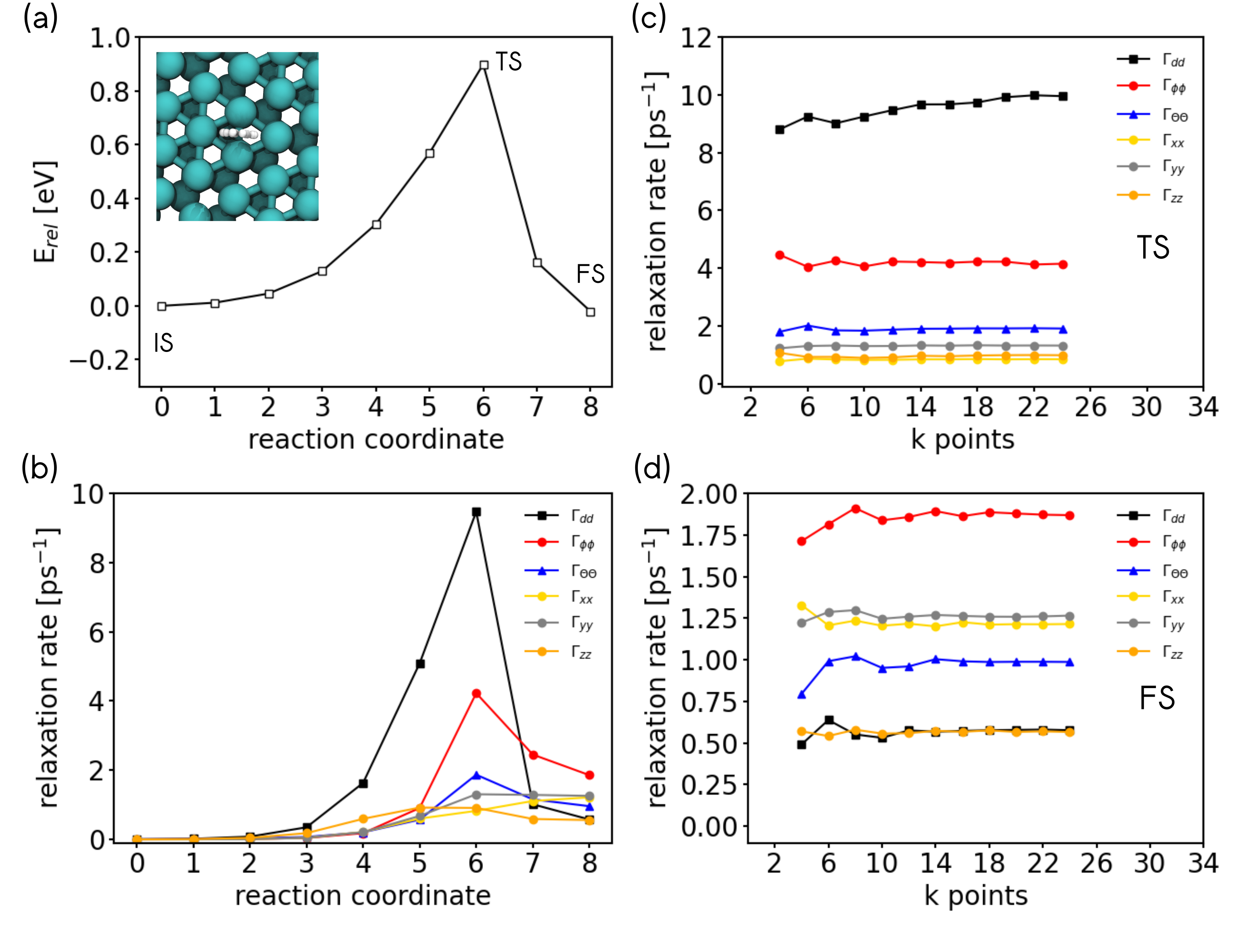}
 \caption{(a) Minimum energy path (MEP) and (b) electronic friction relaxation rates associated with the hydrogen dissociation reaction on a periodic Mg(0001) slab. Initial state, transition state, and final chemisorbed state are abbreviated as IS, TS, and FS, respectively. The panels (c) and (d) show the diagonal elements of the electron friction tensor transformed into molecular internal coordinates for different k-grid values obtained for the transition state (TS) and final state (FS), respectively. The defintion of internal coordinates is given in Ref.~\onlinecite{2021_box}. }
 \label{fgr:zone-all}
\end{figure}

\newpage
\section{projected DOS along minimum energy path}

In order to gain insights into the hydrogen dissociation/recombination reaction, the projected density of states (pDOS) over hydrogen atomic orbitals (AOs) was computed for three different states along the MEP (IS, TS, and FS) for both model systems, for a small Mg nanocluster with 55 metallic atoms and a periodic Mg(0001) slab. Figure S8 shows the pDOS over hydrogen MOs for both systems.

\begin{figure}[!h]
 \centering
 \includegraphics[width=7.5cm]{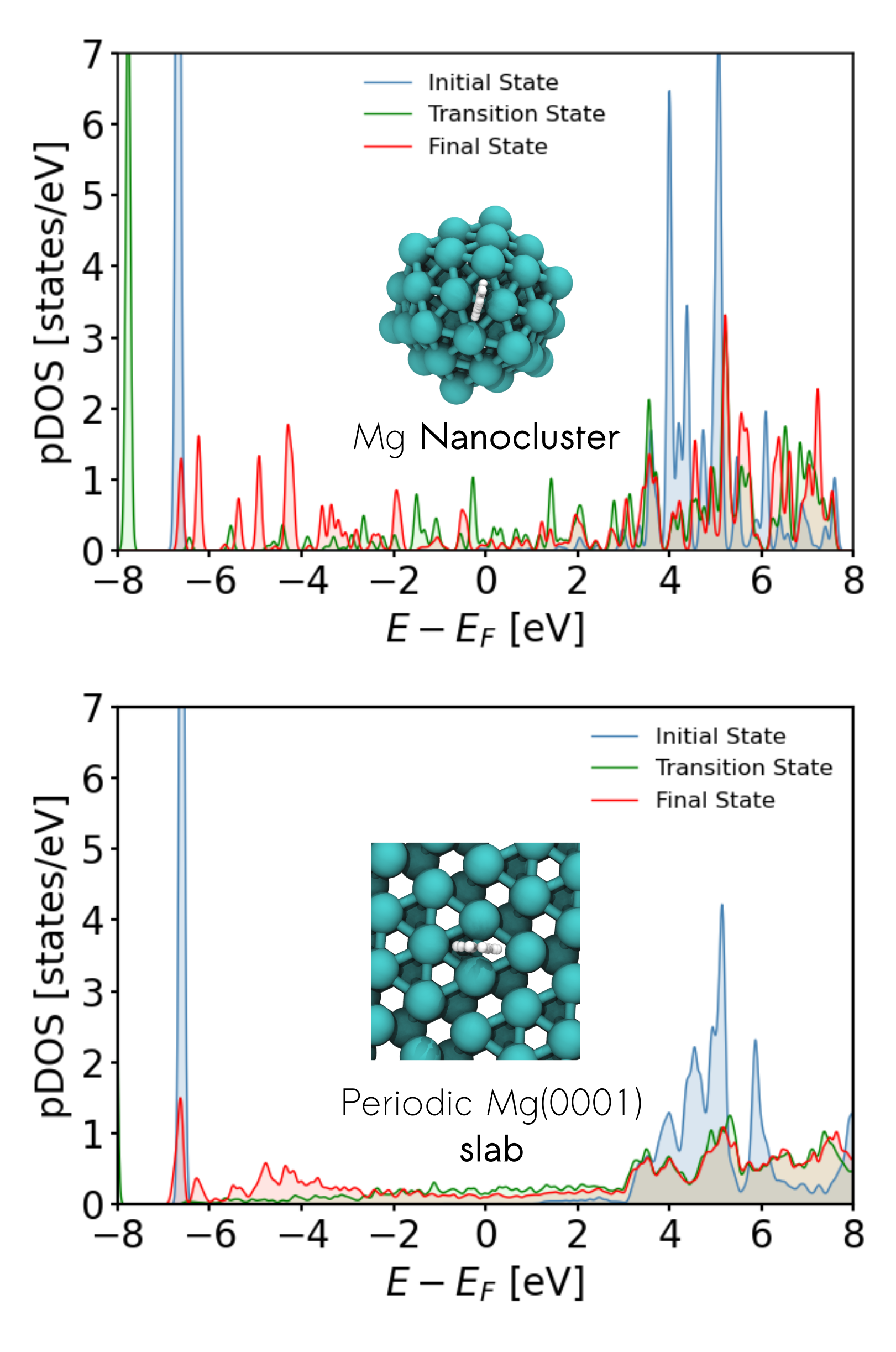}
 \caption{Projected density of state (pDOS) for hydrogen atomic orbitals along the MEP computed for the small Mg nanocluster (top panel) and the periodic Mg(0001) metallic slab at three different states (IS, TS and FS). The pDOS was computed at PBE level by using a similar set up as shown in Figure S1.}
 \label{fgr:zone-all}
\end{figure}\newpage


\section{Two-temperature model}

To describe the electronic ($T_{el}$) and phonon ($T_{ph}$) temperature evolution under ultra-fast laser excitation on a Mg surface, the two temperature model (TTM)\cite{1974_anisimov} was employed. The temporal evolution of both temperatures is described by means of two coupled thermal diffusion equations which in their simplest version (1D-TTM) are:

\begin{equation}
   C_{el} \frac{\partial T_{el}}{\partial t} = \frac{\partial \kappa_{el}} {\partial z}\frac{\partial T_{el}} {\partial z} - G_{0}(T_{el}-T_{ph})+S(z,t) 
\end{equation}

\begin{equation}
   C_{ph} \frac{\partial T_{ph}}{\partial t} = G_{0}(T_{el}-T_{ph})
\end{equation}

Here, C$_{el}$ and C$_{ph}$ are electron and phonon heat capacities, $G_0$ is the \textit{effective electron-phonon-coupling constant} and $\kappa_{el}$ is the electronic heat conductivity. A linear temperature dependence for the electronic heat capacity ($C_{el}= \gamma_{el} T_{el}$) and electronic heat conductivity ($\kappa_{el} = \kappa_{RT} (T_{el}/T_{ph})$) was assumed. 
$S(z,t)$ is the laser source term and $z$ is the vertical position within the metal relative to the surface ($z=0$). A Gaussian function has been selected to describe the laser source term.

The \textit{effective electron-phonon coupling constant} ($G_0$) can be related  with the electron-phonon coupling constant($\lambda$) with following equation,\cite{ep-equation}

\begin{equation}
   G_{0} =\frac{1}{V_c}(\frac{\pi k_B} {\hbar})\times \lambda \langle\omega^2 \rangle \times g(\epsilon_F) 
\end{equation}

where, $V_c$ is the volume of the unit cell, $k_B$ is the Boltzmann's constant, $\hbar$ is the reduced Plack's constant and $g(\epsilon_F)$ is the electronic DOS at the Fermi level. Here, $V_c$ and $g(\epsilon_F) $ values were determined by using our calculated values at PBE level.\cite{2007_lattice-constant}
To compute the $G_{0}$ from the $\lambda$=0.28 value,\cite{mg-lambda} the following approximation was used $\langle\omega^2 \rangle \approx \theta_{D}^{2}/2$. \cite{mg-lambda,zhigilei_2008} Here, $\theta_{D}$ is the Debye temperature for Mg. The final computed value is $G_{0} = 6.75\times 10^{16}$ W K$^{-1}$ m$^{-3}$.  The other used thermophysical properties were, $\gamma_{el}$ = 71.00 J m$^{-3}$ K$^{-2}$ (electron specific heat constant), $\kappa_{RT}$ = 156.00 W m$^{-1}$ K$^{-1}$  (thermal conductivity at 300K), $\theta_{D}$ = 400 K (Debye Temperature).\cite{kittel_book} For the laser source, a Gaussian form was selected  similar to Ref. \onlinecite{saalfrank_2019} (FWHM = 150 nm and $\tau$ = $\frac{FWHM}{2 \sqrt{2ln2}}$ and optical penetration depth $\zeta$ = 8.96 nm)

\section{Laser-driven heating of hydrogen atoms on Mg}

We estimate the kinetic energy that can be transferred via hot electrons into chemisorbed hydrogen atoms by propagating an Ornstein-Uhlenbeck process for each of the three degrees of freedom of the atom:~\cite{Leimkuhler2013}
\begin{equation}\label{eq:OU}
dv= -\gamma v dt + \sqrt{k_B T M^{-1}2\gamma}dW 
\end{equation}
The first term on the right hand side corresponds to the friction contribution and the second term to the random force that describes energy transfer from the bath of hot electrons to the adsorbate. $\gamma$ is the friction coefficient, $T$ is the electronic temperature, and $dW$ is a random Wiener process.

According to the relaxation rates we find in Figure 5 of the main manuscript, we assume a friction coefficient that corresponds to a relaxation rate of 2~ps$^{-1}$. The three temperature profiles calculated with the TTM and shown in the inset of Figure 5 of the main manuscript are used to drive the time-dependent electronic temperature in eq.~\ref{eq:OU}. They correspond to laser fluences of 10, 50, and 100~J/m$^2$ leading to peak electronic temperatures of 2500, 5500, and 8000~K, respectively. For each condition, we simulate 1000 trajectories and calculate the average kinetic energy of hydrogen at 250~fs after $t=0$ of the laser pulse. For 10, 50, and 100~J/m$^2$, we find a kinetic energy of the H atom of 0.17, 0.40, and 0.55~eV.
\newpage

\section{Full-optimization effect on Mg plasmonic response}

\begin{figure}
 \centering
 \includegraphics[width=8cm]{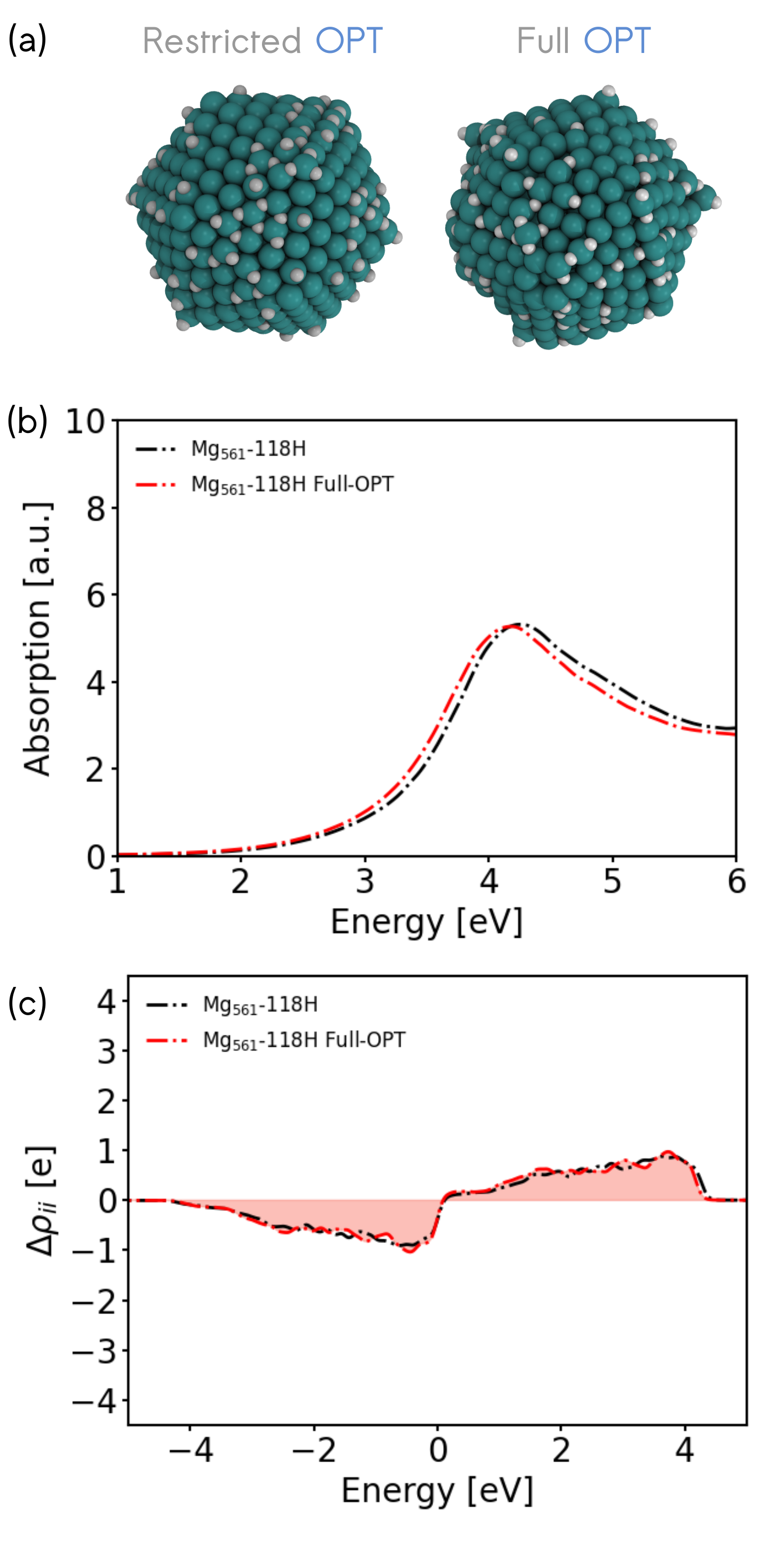}
 \caption{(a) Optimized structures for a Mg nanocluster with 118 H atoms with and without surface relaxation. (b) Optical absorption response and (c) hot-carrier distribution for both structures consider in (a). The  electric field intensity  used to produce the hot-carrier distribution in both configuration was $E_0$=0.02~V~$\AA^{-1}$. In this figure, "OPT" is referring to structural optimization.}
 \label{fgr:zone-all}
\end{figure}

The full optimization process generates a local surface relaxation on the Mg metal surface which has only minor effects on the optical absorption properties and hot-carrier distribution. The potential effects associated with the plasmonic response may be harnessed while the plasmonic properties and metallic character are held which is what is expected to happen during the early hydrogen adsorption stage. 


%